\def\be{\begin{equation}}
\def\ee{\end{equation}}
\def\ba{\begin{array}}
\def\ea{\end{array}}

\documentclass[prl,showpacs,twocolumn,amsmath]{revtex4}
\usepackage{amsfonts}
\usepackage{mathrsfs}
\usepackage{mathdots}
\usepackage{amssymb}
\usepackage{pifont}
\usepackage{epsfig,subfigure,dsfont,amsthm,amsbsy,mathrsfs,amscd}

\UseRawInputEncoding
\usepackage{rotating}

\usepackage{graphicx}
\usepackage{dcolumn}
\usepackage{bm}
\usepackage{amsmath}
\usepackage{multirow}
\usepackage{array}
\usepackage{epstopdf}
\usepackage{diagbox}
\usepackage{dcolumn}

\def\qed{\leavevmode\unskip\penalty9999 \hbox{}\nobreak\hfill
     \quad\hbox{\leavevmode  \hbox to.77778em{%
               \hfil\vrule   \vbox to.675em%
               {\hrule width.6em\vfil\hrule}\vrule\hfil}}
     \par\vskip3pt}

\begin{document}

\title{Gaussian quantum steering of coupled three-mode squeezed vacuum in an expanding universe}
\author{Guang-Wei Mi}
\author{Xiaofen Huang}
\author{Tinggui Zhang$^{*}$}
\affiliation{School of Mathematics and Statistics, Hainan Normal University, Haikou, 571158, China\\
{$^{*}$ Email address: tinggui333@163.com}}


\begin{abstract}
The coupled three-mode squeezed vacuum is a representative multimode squeezed Gaussian state featuring unique steerability. This work investigates Gaussian quantum steering distributions of the coupled three-mode squeezed vacuum under an expanding universe. Due to causal separation between the interior and exterior spacetime regions, quantum information behind the event horizon is inaccessible to Alice, Bob and Charlie. We separately analyze steering behaviors for physically accessible and inaccessible modes. Our analysis shows that greater total mean photon number and momentum, combined with reduced expansion volume and expansion rate, enhance quantum steering strength. Notably, Gaussian quantum steering for physically inaccessible modes undergoes the ``sudden death" phenomenon when a critical threshold parameter $\phi$ is exceeded. These results deliver novel insights into quantum correlations in curved spacetime.
\end{abstract}

\maketitle


\section{Introduction}
Quantum correlations have been extensively investigated in recent years, manifesting themselves in various forms including entanglement, quantum steering, and Bell nonlocality ~\cite{Modi.2012}. Originally proposed by Schr\"{o}dinger back in 1935 as a generalization of the Einstein-Podolsky-Rosen (EPR) paradox, quantum steering describes the ability for one party to remotely alter the quantum states of a spatially separated system through local measurement operations~\cite{Einstein.1935,Schrodinger.1935,Schrodinger.1936}.
Since Wiseman \emph{et al.}~\cite{Wiseman.2007} put forward the formal definition of quantum steering via the local hidden state (LHS) model in 2007 and clarified its operational rank as an intermediate quantum correlation situated between entanglement and Bell nonlocality, the subject has drawn revitalized research focus~\cite{Jones.2007,Walborn.2011,Navascues.2012,Skrzypczyk.2014,Deng2017,Nery.2018,Designolle.2021}. Moreover, the asymmetry inherent to quantum steering within quantum systems has been experimentally verified~\cite{Saunders.2010,Wollmann.2016,Xiao.2017}. These distinctive traits of quantum steering carry profound implications for quantum communication protocols including quantum key distribution (QKD), quantum teleportation, remote quantum state preparation, and dense coding~\cite{Branciard.2012,Reid.2013,Piani.2015,Tischler.2018,Fan.2022}, which boost both the efficiency and security of such schemes.

The two-mode squeezed vacuum (TMSV) arguably constitutes the most prevalent EPR-entangled quantum resource~\cite{Horodecki.2009}.  Driven by advancing quantum technologies and their practical demands, an ever-expanding library of entangled systems has been proposed and deployed over recent years~\cite{Loock.2000,Aoki.2003,Gonzalez.2018,Xie.2021,Suprano.2022}. Zhang and Glasser~\cite{Zhang.2022} proposed the coupled three-mode squeezed vacuum (C3MSV), a quantum state that hosts genuine tripartite entanglement.

Motivated by the endeavor to unify quantum mechanics and general relativity, studies concerning quantum information within noninertial reference frames and curved spacetime have grown into a fast-expanding research frontier in recent years~\cite{Alsing.2003,Schuller.2005,Pan.2008,Wang.2009,Esfahani.2011,Dong.2018,Das.2019,Bhattacharya.2022,Wu.2022,Li.2022,Zhang.2023,Mi.2025,Wu.2025,Mi.2026}.
In~\cite{Wang.2016}, Wang \emph{et al.} investigated the Gaussian quantum steering and its asymmetry in the Schwarzschild black hole. In~\cite{Zhan.2023}, Zhan \emph{et al.} explored the Gaussian steering and remote generation of Wigner negativity for the C3MSV. In~\cite{PLB.2025}, Wu \emph{et al.} study the Gaussian tripartite steering in Schwarzschild black hole. In~\cite{EPJC.2024}, Wu \emph{et al.} investigated Gaussian quantum steering for continuous variables sharing in an expanding universe.
Moreover, growing research focus has been devoted to the dynamical control of quantum systems subjected to dynamical Casimir effects~\cite{Plunien.1986}, Hawking radiation~\cite{Hawking.1975}, and relativistic kinematics. Hence, it is of great importance to investigate quantum steering signatures within relativistic formalisms and expanding universe.

In the paper, we investigate the distribution of the Gaussian quantum steering for the C3MSV in an expanding spacetime. It is postulated that Alice, Bob and Charlie initially share a C3MSV state in asymptotically flat spacetime. After that, Charlie enters a spacetime affected by cosmic expansion, with Alice and Bob remaining in the flat asymptotic domain. Causal disconnection between the inner and outer regions of the expanding spacetime prevents all three parties from obtaining quantum information beyond the event horizon. Then, we investigate Gaussian steering quantities for physically accessible and inaccessible modes.
For Gaussian quantum steering associated with both physically accessible and inaccessible modes, our analysis reveals that increased total mean photon number and momentum, together with diminished expansion volume and expansion rate, give rise to enhanced steering.
In the case of inaccessible modes, ``sudden death" phenomenon emerges within some steering configurations, an effect that never occurs for accessible modes independently.

The rest of this paper is organized as follows. In Sec.\uppercase\expandafter{\romannumeral2}, we briefly outline the cosmic expansion process modeled via Gaussian channels. In Sec.\uppercase\expandafter{\romannumeral3}, we  introduce the definitions and quantification criteria of Gaussian states and Gaussian quantum steering.
In Sec. \uppercase\expandafter{\romannumeral4}, we investigate distribution of C3MSV Gaussian steering in an expanding spacetime.
We conclude in Sec.\uppercase\expandafter{\romannumeral5}.

\section{Gaussian-channel-based description of cosmic expansion}

We commence our analysis in a $1+1$-dimensional Robertson-Walker expanding spacetime, whose metric takes the form
\begin{eqnarray}
\begin{aligned}
ds^{2}=dt^{2}-[a(t)]^{2}dx^{2},
\end{aligned}
\end{eqnarray}
where $a(t)$ denotes the scale factor. Introducing conformal time $\eta$, which is connected to the cosmological time $t$ via the integral definition
$\eta=\int_{0}^{t}\frac{d\tau}{a(\tau)}$, one may recast the metric of the Robertson-Walker expanding universe as~\cite{Ball.2006,Fuentes.2010,Martinez.2012}
\begin{eqnarray}
\begin{aligned}
ds^{2}=[a(\eta)]^{2}(d\eta^{2}-dx^{2}).
\end{aligned}
\end{eqnarray}
Within this framework, the conformal scale factor reads
\begin{eqnarray}
\begin{aligned}
&&[a(\eta)]^{2}=1+\epsilon(1+\tanh(\sigma\eta)),
\end{aligned}
\end{eqnarray}
where the parameters $\epsilon$ and $\sigma$ describe the volume and the rapidity of cosmic expansion, respectively. One can readily verify that the spacetime geometry reduces to flat Minkowski spacetime in the distant past and far future: the metric simplifies to $ds^2=d\eta^2-dx^2$ for $\eta\to-\infty$, while it takes the form $ds^2=(1+2\epsilon)(d\eta^2-dx^2)$ as $\eta\to+\infty$. For this reason, the timelike Killing vector field and the particle spectrum of the quantum field are well-defined within these two limits.

Then, consider a real scalar field $\phi(x,\eta)$ residing in the Robertson-Walker expanding spacetime, which satisfies the Klein-Gordon equation
\begin{eqnarray}
\begin{aligned}
(\square+m^2)\Phi=0,
\end{aligned}
\end{eqnarray}
where $\square=\frac{1}{\sqrt{|g|}}\partial_\mu\sqrt{|g|}g^{\mu\nu}\partial_\nu$. Upon solving the Klein-Gordon equation in the two limits $\eta\to\pm\infty$, we construct the orthonormal basis of $u^\text{in}$ modes valid over the distant past $(``\text{in}")$ domain and its complementary counterpart $u^\text{out}$ mode basis for the far future $(``\text{out}")$ domain. Making use of the inner product, the Bogoliubov transformation relating the modes $u_k^\text{in}$ and $u_k^\text{out}$ can be expressed as
\begin{eqnarray}
\begin{aligned}
u_k^\text{in}(x,\eta)=\alpha_k u_k^\text{out}(x,\eta)+\beta_k u_{-k}^{\text{out}*}(x,\eta),
\end{aligned}
\end{eqnarray}
where the Bogoliubov coefficients take the explicit form
\begin{eqnarray}
\begin{aligned}
&\alpha_k=\sqrt{\frac{\omega_\text{out}}{\omega_\text{in}}}\frac{\Gamma\left(1-\frac{i\omega_\text{in}}{\sigma}\right)\Gamma\left(-\frac{i\omega_\text{out}}{\sigma}\right)}
{\Gamma\left(1-\frac{i\omega_+}{\sigma}\right)\Gamma\left(-\frac{i\omega_+}{\sigma}\right)},\\
&\beta_k=\sqrt{\frac{\omega_\text{out}}{\omega_\text{in}}}\frac{\Gamma\left(1-\frac{i\omega_\text{in}}{\sigma}\right)\Gamma\left(\frac{i\omega_\text{out}}{\sigma}\right)}
{\Gamma\left(1+\frac{i\omega_-}{\sigma}\right)\Gamma\left(\frac{i\omega_-}{\sigma}\right)}.
\end{aligned}
\end{eqnarray}
Here, $\Gamma$ denotes the gamma function, $\omega_\text{in}=\sqrt{k^2+m^2}$, $\omega_\text{out}=\sqrt{k^2+m^2(1+2\epsilon)}$, $\omega_\pm=\frac{1}{2}(\omega_\text{out}\pm \omega_\text{in})$. Elementary algebraic manipulation verifies that the Bogoliubov coefficients obey $|\alpha_k|^2-|\beta_k|^2=1$. To streamline subsequent derivations, we introduce $\theta_k^2 = \left|\frac{\beta_k}{\alpha_k}\right|^2 = \frac{\sinh^2\left(\pi\frac{\omega_-}{\sigma}\right)}{\sinh^2\left(\pi\frac{\omega_+}{\sigma}\right)}$, from which we immediately deduce
\begin{eqnarray}
\begin{aligned}
|\alpha_k|^2=\frac{1}{1-\theta_k^2},\quad |\beta_k|^2=\frac{\theta_k^2}{1-\theta_k^2},
\end{aligned}
\end{eqnarray}
where $|\beta_k|^2$ corresponds to the mean particle occupation number generated in the ``out" mode $k$. Accordingly, the limit $\theta_k^2 \to 0$ implies the average particle count of mode $k$ vanishes, whereas $\theta_k^2 \to 1$ signifies that the mean particle number for mode $k$ diverges to infinity.

The bosonic annihilation and creation operators obey the Bogoliubov linear mapping relations
\begin{eqnarray}
\label{bin}
\begin{aligned}
b_{\text{in},k}=\alpha_k^* b_{\text{out},k} - \beta_k^* b_{\text{out},-k}^\dagger,
\end{aligned}
\end{eqnarray}
\begin{eqnarray}
\begin{aligned}
b_{\text{in},k}^\dagger=\alpha_k b_{\text{out},k}^\dagger - \beta_k b_{\text{out},-k}.
\end{aligned}
\end{eqnarray}
Here, $b_{\text{in},k}$ and $b_{\text{in},k}^\dagger$ denote the annihilation and creation operators for bosons acting on the Hilbert space of states in the asymptotic past regime, $b_{\text{out},k}$ and $b_{\text{out},k}^\dagger$ correspond to their counterparts defined on the asymptotic future states, and $b_{\text{out},-k}$ and $b_{\text{out},-k}^\dagger$ act as the annihilation and creation operators for antiparticles, respectively.
We exploit the defining property of $b_{\text{in},k}|0_k\rangle_\text{in}=0$ to derive the correspondence between the ``in" and ``out" vacuum states.
Substituting Eq.(\ref{bin}) for $b_{\text{in},k}$ into this vacuum condition yields
\begin{eqnarray}
\begin{aligned}
\left(\alpha_k^* b_{\text{out},k}-\beta_k^* b_{\text{out},-k}^\dagger\right)|0_k\rangle_\text{in}=0.
\end{aligned}
\end{eqnarray}
By virtue of the Bogoliubov normalization constraint, the ``in" vacuum state can be written in the asymptotic future as
\begin{eqnarray}
\begin{aligned}
|0_k\rangle_\text{in}=\sum_{n=0}^{\infty} A_n |n_k\rangle_\text{out}|n_{-k}\rangle_\text{out},
\end{aligned}
\end{eqnarray}
where $A_n=\sqrt{1-\theta_k^2}\left(\frac{\beta_k^*}{\alpha_k}\right)^n$, $n_k$ represents the boson number, and $n_{-k}$ represents antiboson number. This decomposition demonstrates that the primordial vacuum state $|0_k\rangle_\text{in}$ evolves into a canonical two-mode squeezed state within the far future. Upon rotating the squeezing parameter and absorbing the global phase factor, we arrive at~\cite{Caves.1985,Schumaker.1985,Adesso2007}
\begin{eqnarray}
\label{0k}
\small
\begin{aligned}
|0_k\rangle_\text{in}=\sqrt{1-\theta_k^2}\sum_{n=0}^{\infty} \theta_k^n |n_k\rangle_\text{out}|n_{-k}\rangle_\text{out}=U_k |0_k\rangle|0_{-k}\rangle.
\end{aligned}
\end{eqnarray}
Here, $U_k=\exp\left[r_k\left(b_{\text{out},k}^\dagger b_{\text{out},-k}^\dagger-b_{\text{out},k} b_{\text{out},-k}\right)\right]$ stands for a two-mode squeezing operator,
where the squeezing parameter $r_k$ is fixed by the algebraic relation $\cosh(r_k)=|\alpha_k|$.
It is imperative to emphasize that $U_k$ implements a Gaussian quantum operation, which preserves the Gaussian statistical profile of all input quantum states. Hence, Eq.(\ref{0k}) demonstrates that the cosmological expansion of the Robertson-Walker spacetime can be modeled as a Gaussian channel equivalent to a bosonic amplification process.
Within the phase-space formulation of quantum mechanics, the action induced by $U_k$ corresponds to a symplectic transformation characterized by the matrix
\begin{eqnarray}
\label{symplectic}
S_k=\frac{1}{\sqrt{1-\theta_k^2}}
\left (
\begin{array}{cc}
\mathbb{I}_2       &\theta_k Z_2\\
\theta_k Z_2       &\mathbb{I}_2
\end{array}
\right ),
\end{eqnarray}
where $\mathbb{I}_{2}$ is the $2\times2$ identity matrix and $Z_2$ is the third Pauli matrix.

\section{Gaussian states and Gaussian quantum steering}

First, we briefly review Gaussian states, which occupy a prominent role in the field of continuous variable quantum technologies~\cite{Weedbrook.2012}. In the paper, we consider a continuous variable quantum system consisting of $(n+m)$ bosonic modes in the bipartite state $\rho_{AB}$, where subsystem $A$ contains $n$ modes and subsystem $B$ contains $m$ modes, respectively. The quadrature components of displacement and momentum for each mode can be assembled into a column vector
\begin{eqnarray}
\begin{aligned}
\label{R}
\hat{R}&=(\hat{x}^{A}_{1}, \hat{p}^{A}_{1},\ldots, {x}^{A}_{n}, \hat{p}^{A}_{n}, \hat{x}^{B}_{1}, \hat{p}^{B}_{1},\ldots, {x}^{B}_{m}, \hat{p}^{B}_{m})^{T}\\
&=(\hat{R}_{1}, \hat{R}_{2},\ldots, \hat{R}_{2(n+m)-1}, \hat{R}_{2(n+m)})^{T}.
\end{aligned}
\end{eqnarray}
Here, $\hat{x}_{k}^{A(B)}$ and $\hat{p}_{k}^{A(B)}$ denote the position and momentum operators associated with the $k$-th mode of the subsystem $A(B)$, respectively $(k=1,2,\ldots,n(m))$. These operators are expressed in terms of the creation and annihilation operators of the mode
\begin{eqnarray*}
\begin{aligned}
\hat{x}_{k}=\hat{a}_{k}+\hat{a}^{\dag}_{k}, \quad \quad \hat{p}_{k}=-i(\hat{a}_{k}-\hat{a}^{\dag}_{k}).
\end{aligned}
\end{eqnarray*}
In addition, Eq.(\ref{R}) satisfies the canonical commutation relations $[\hat{R}_{i},\hat{R}_{j}]=2i\delta_{ij}$, with $\delta=\bigoplus_{1}^{n+m}\begin{pmatrix} 0 & 1 \\ -1 & 0 \end{pmatrix}$ being the symplectic form.
Every Gaussian state $\rho_{AB}$ is uniquely determined by its first and second statistical moments. The displacement vector $\bar{d}=(d_1, d_2, \dots, d_{2(n+m)})^\mathrm{T}$ associated with $\rho_{AB}$ is formulated as~\cite{Braunstein.2005}
\begin{eqnarray}
\begin{aligned}
\bar{d}&=\big(\langle\hat{R}_1\rangle,\langle\hat{R}_2\rangle,\dots,\langle\hat{R}_{2(n+m)}\rangle\big)^\mathrm{T} \\
&=\big(\mathrm{Tr}(\rho\hat{R}_1), \mathrm{Tr}(\rho\hat{R}_2), \dots, \mathrm{Tr}(\rho\hat{R}_{2(n+m)})\big)^\mathrm{T},
\end{aligned}
\end{eqnarray}
while the covariance matrix $(CM)$ $\sigma=(\sigma_{ij})$ is specified via~\cite{Braunstein.2005}
\begin{eqnarray}
\begin{aligned}
\sigma_{ij}&=\frac{1}{2}Tr\left[ \{\hat{R}_{i},\hat{R}_{j}\}_{+} \rho_{AB}\right]\\
&=\frac{1}{2}\langle\hat{R}_{i}\hat{R}_{j}+\hat{R}_{j}\hat{R}_{i}\rangle-\langle\hat{R}_{i}\rangle\langle\hat{R}_{j}\rangle.
\end{aligned}
\end{eqnarray}
The $CM$ can be cast into a block form
\begin{eqnarray}
\sigma_{AB}=
\left (
\begin{array}{cc}
A        &C              \\
C^{T}    &B       \\
\end{array}
\right ).
\end{eqnarray}
The CM $\sigma_{AB}$ corresponds to a valid quantum state if and only if it obeys the physical uncertainty principle constraint
\begin{eqnarray}
\begin{aligned}
\sigma_{AB}+i(\delta_{AB})\geq 0.
\end{aligned}
\end{eqnarray}

Next, we introduce the concept of quantum steering. Given that Alice carries out a set of measurements $\mathcal{M}_A$, the bipartite state $\rho_{AB}$ is $A\to B$ steering iff the joint probability $P(r_A,r_B|R_A,R_B,\rho_{AB})=\sum_\lambda \wp_\lambda \wp(r_A|R_A,\lambda)P(r_B|R_B,\rho_\lambda)$ cannot hold for every pair of local observables $R_A$ (on A, outcome $r_A$) and $R_B$~\cite{Wiseman.2007}.
In short, one pair $(R_A,R_B)$ must invalidate this expression with a globally fixed $\wp_\lambda$. Here, $\wp_\lambda$ and $\wp(r_A|R_A,\lambda)$ are probability distributions, and $P(r_B|R_B,\rho_\lambda)$ is the conditional probability for the hidden-state ensemble $\rho_\lambda$.
It was demonstrated in~\cite{Wiseman.2007} that a general $(n+m)$-mode Gaussian state $\rho_{AB}$ is $A \to B$ steerable under Alice’s Gaussian measurements iff
\begin{eqnarray}
\begin{aligned}
\label{sigmaAB}
\sigma_{AB} + i(0_A \oplus \delta_B) \geq 0
\end{aligned}
\end{eqnarray}
is violated. Thus, breaking inequality (\ref{sigmaAB}) is necessary and sufficient for Gaussian $A \to B$ steering.

Then, the steering from Alice to Bob via Gaussian measurements can be quantified by~\cite{Kogias.2015}
\begin{eqnarray}
\begin{aligned}
\label{m=2}
\mathcal{G}^{A\to B}(\sigma_{AB}) := \max\left\{0,\,-\sum_{j:\bar{\nu}_j^B<1}\ln(\bar{\nu}_j^B)\right\},
\end{aligned}
\end{eqnarray}
where the set $\{\bar{\nu}_j^B\}$ denotes the symplectic eigenvalues of the Schur complement of subsystem A within the covariance matrix $\sigma_{AB}$.
$\mathcal{G}^{A\to B}$ is monotonic with respect to Gaussian local operations and classical communication~\cite{Lami.2016}, and equals zero if Alice cannot achieve $A\to B$ steering via Gaussian measurements. This measure has been experimentally demonstrated with Gaussian cluster states through covariance matrix reconstruction~\cite{Deng2017}.
The Gaussian $A\to B$ steering measure takes an extremely straightforward expression when Bob holds only one mode $(i.e., m=1)$~\cite{Kogias.2015}
\begin{eqnarray}
\label{m=1}
\begin{aligned}
\mathcal{G}^{A\to B}(\sigma_{AB}) &= \max\left\{0,\,\frac12\ln\frac{\det A}{\det \sigma_{AB}}\right\} \\
&= \max\left\{0,\, \mathcal{S}(A)-\mathcal{S}(\sigma_{AB})\right\},
\end{aligned}
\end{eqnarray}
where $\mathcal{S}(\sigma)=\frac12\ln(\det\sigma)$ defines the R$\acute{e}$nyi-2 entropy ~\cite{Adesso.2012}.
Similarly, one acquires the Gaussian $B\to A$ steering measure by swapping subsystems A and B.

It is well established that EPR steering constitutes a directional manifestation of quantum nonlocality and inherently features asymmetric characteristics. In practical implementations, quantum steering is classified into three mutually exclusive regimes: (i) no-way steering, featuring vanishing steerability along all directional channels;(ii) one-way steering, where the state manifests steerability restricted to a single orientation only; (iii)  two-way steering, in which the shared quantum state admits steerability over both inverse directional pathways.

\section{Distribution of C3MSV Gaussian Steering in an Expanding Spacetime}

Firstly, we briefly review C3MSV, a pivotal resource within continuous variable quantum information theory.
The three-mode interaction Hamiltonian $H=i\hbar\left(\eta_1^* a_1 a_2 + \eta_2^* a_2 a_3 - \eta_1 a_1^\dagger a_2^\dagger - \eta_2 a_2^\dagger a_3^\dagger\right)$
is physically implementable via a doubly pumped four-wave mixing (FWM) scheme~\cite{Zhang.2022}.
This Hamiltonian generates the unitary time evolution operator dubbed C3MSO~\cite{Zhan.2023}
\begin{eqnarray}
\begin{aligned}
S_3 = e^{\xi_1^* a_1 a_2 + \xi_2^* a_2 a_3 - \xi_1 a_1^\dagger a_2^\dagger - \xi_2 a_2^\dagger a_3^\dagger},
\end{aligned}
\end{eqnarray}
where $\xi_j = \eta_j t = r_j e^{i\theta_j} (j=1,2)$ denote two complex squeezing parameters with respective magnitudes $r_j$ and phases $\theta_j$.
To simplify subsequent calculations, we reparameterize the pair $(r_1,r_2)$ into polar coordinates $(r,\phi)$ defined by
$r=\sqrt{r_1^2+r_2^2},\quad \cos\phi=\frac{r_1}{r},\quad \sin\phi=\frac{r_2}{r}$, where $\phi\in[0,\frac{\pi}{2}]$.

The three-mode squeezing operator $S_3$ acts on the product state of three uncorrelated vacuum modes $|000\rangle$ to generate the C3MSV state given by
\begin{eqnarray}
\begin{aligned}
|\psi\rangle \equiv S_3|000\rangle = \frac{1}{c}e^{-\frac{\epsilon_1}{c}a_1^\dagger a_2^\dagger-\frac{\epsilon_2}{c}a_2^\dagger a_3^\dagger}|000\rangle,
\end{aligned}
\end{eqnarray}
where $c=\cosh r$, $s=\sinh r$, $\epsilon_1 = s e^{i\theta_1}\cos\phi$, and $\epsilon_2 = s e^{i\theta_2}\sin\phi$.

The CM characterizing the C3MSV state takes the form~\cite{Zhan.2023,Weedbrook.2012,Genoni.2016,Brandao.2022}
\begin{eqnarray}
\label{CM}
V=
\left (
\begin{array}{ccc}
(1+2\bar{n}_1)\mathbb{I}_2                       & -2sc\cos\phi\,\Sigma_{\theta_1}      & s^2\sin2\phi\,R_{\theta_2-\theta_1} \\
-2sc\cos\phi\,\Sigma_{\theta_1}                  & (1+2\bar{n}_2)\mathbb{I}_2           & -2sc\sin\phi\,\Sigma_{\theta_2} \\
s^2\sin2\phi\,R_{\theta_2-\theta_1}      & -2sc\sin\phi\,\Sigma_{\theta_2}      & (1+2\bar{n}_3)\mathbb{I}_2
\end{array}
\right ),
\end{eqnarray}
Here, $\bar{n}_1 = s^2\cos^2\phi,\quad \bar{n}_2 = s^2,\quad \bar{n}_3 = s^2\sin^2\phi$, $\mathbb{I}_2$ denotes the $2\times2$ identity matrix, and

$\Sigma_\theta=
\begin{pmatrix}
\cos\theta & \sin\theta \\
\sin\theta & -\cos\theta
\end{pmatrix},\quad
R_\theta=
\begin{pmatrix}
\cos\theta & \sin\theta \\
-\sin\theta & \cos\theta
\end{pmatrix}.$

Notably, for subsequent numerical simulations, we commonly reparameterize the squeezing parameter $r$ in terms of the total mean photon number $\bar{n}_T=2s^2$ via the relation $r=\mathrm{arcsinh}\sqrt{\bar{n}_T/2}$, and fix the phase angles $\theta_1=\theta_2=0$.

Then, we assume that Alice, Bob and Charlie initially share the C3MSV state $\sigma_{ABC}^{\text{in}}=V$ within an asymptotically flat spacetime. Next, Charlie undergoes cosmic expansion governed by the symplectic transformation specified in Eq.(\ref{symplectic}), while Alice and Bob remain situated in the asymptotically flat region. Consequently, the initial state $\sigma_{ABC}^{\text{in}}$ evolves into a four-mode Gaussian state in the asymptotic future. In other words, the primordial C3MSV defined in the asymptotic past is mapped to a four-mode Gaussian state under the expansion of the Robertson-Walker spacetime as time approaches infinity. Accordingly, the full quantum system comprises four bosonic modes: mode A held by Alice, mode B held by Bob, mode C held by Charlie, and mode $\bar{C}$ held by anti-Charlie.
Due to causal isolation between the interior and exterior regions of the expanding spacetime, Alice, Bob, and Charlie cannot retrieve quantum information behind the event horizon. We therefore classify the fermionic modes $(A, B, C)$ as accessible and the antifermionic mode $\bar{C}$ as inaccessible.

The complete covariance matrix characterizing the complete system is therefore given by~\cite{Adesso2007}
\begin{eqnarray}
\label{ABCc}
\begin{aligned}
\sigma_{ABC\bar{C}}^{\text{out}} &= \big[I_{AB}\oplus S_{C,\bar{C}}\big]\big[\sigma_{ABC}^{in}\oplus I_{\bar{C}}\big]
\big[I_{AB}\oplus S_{C,\bar{C}}\big]^T \\
&=
\begin{pmatrix}
\sigma_A                         & \mathcal{E}_{AB}                & \mathcal{E}_{AC}             & \mathcal{E}_{A\bar{C}} \\
\mathcal{E}_{AB}^T               & \sigma_B                        & \mathcal{E}_{BC}             & \mathcal{E}_{B\bar{C}} \\
\mathcal{E}_{AC}^T               & \mathcal{E}_{BC}^T              & \sigma_C                     & \mathcal{E}_{C\bar{C}} \\
\mathcal{E}_{A\bar{C}}^T         & \mathcal{E}_{B\bar{C}}^T        & \mathcal{E}_{C\bar{C}}^T     & \sigma_{\bar{C}}
\end{pmatrix}.
\end{aligned}
\end{eqnarray}
Here, $S_{C,\bar{C}}$, defined in Eq.(\ref{symplectic}), serves as the phase-space representation of the two-mode squeezing operation, while $I_{AB}$ denotes the $4\times4$ identity matrix and $I_{\bar{C}}$ the $2\times2$ identity matrix. The diagonal elements of the matrix in Eq.(\ref{ABCc}) adopt the following explicit forms
\begin{eqnarray}
\begin{aligned}
&\sigma_A=(1+2s^2\cos^2\phi)\mathbb{I}_2,\\
&\sigma_B=(1+2s^2)\mathbb{I}_2,\\
&\sigma_C=\frac{1}{1-\theta_k^2}(\theta_k^2+1+2s^2\sin^2\phi)\mathbb{I}_2,\\
&\sigma_{\bar{C}}=\frac{1}{1-\theta_k^2}[1+\theta_k^2(1+2s^2\sin^2\phi)]\mathbb{I}_2.\\
\end{aligned}
\end{eqnarray}
The off-diagonal elements adopt the forms outlined below
\begin{eqnarray}
\begin{aligned}
&\mathcal{E}_{AB}=-2sc\cos\phi Z_2,\\
&\mathcal{E}_{AC}=\frac{s^{2}\sin2\phi}{\sqrt{1-\theta_k^2}}\mathbb{I}_2,\\
&\mathcal{E}_{A\bar{C}}=\frac{\theta_k s^{2}\sin2\phi}{\sqrt{1-\theta_k^2}}Z_2,\\
&\mathcal{E}_{BC}=\frac{-2sc\sin\phi}{\sqrt{1-\theta_k^2}}Z_2,\\
&\mathcal{E}_{B\bar{C}}=\frac{-2\theta_k sc\sin\phi}{\sqrt{1-\theta_k^2}}\mathbb{I}_2,\\
&\mathcal{E}_{C\bar{C}}=\frac{2\theta_k}{1-\theta_k^2}(1+s^2\sin^2\phi)Z_2.\\
\end{aligned}
\end{eqnarray}

\subsection{A. Gaussian steering distribution of physically accessible modes}

In this subsection, we investigate the Gaussian steering distribution for physically accessible modes.
Then, the covariance matrix governing the three physically accessible modes A, B, and C is derived by tracing out the physically inaccessible mode $\bar{C}$, and takes the form
\begin{eqnarray}
\label{ABC}
\begin{aligned}
\sigma_{ABC}^{\text{out}}=
\begin{pmatrix}
\sigma_A                         & \mathcal{E}_{AB}                & \mathcal{E}_{AC}             \\
\mathcal{E}_{AB}^T               & \sigma_B                        & \mathcal{E}_{BC}             \\
\mathcal{E}_{AC}^T               & \mathcal{E}_{BC}^T              & \sigma_C                     \\
\end{pmatrix}.
\end{aligned}
\end{eqnarray}

Employing the covariance matrix $\sigma_{ABC}^{\text{out}}$ and Eq.(\ref{m=1}), we deduce the analytic expressions for Gaussian  steering corresponding to all $2\rightarrow1$ partitions of the quantum state defined in Eq.(\ref{ABC}). The resulting steering metrics read as
\begin{eqnarray}
\begin{aligned}
&\mathcal{G}^{BC\rightarrow A}(\sigma_{ABC}^{\text{out}})=\max\left\{0,\, \frac{1}{2}\ln \frac{\det BC}{\det\sigma_{ABC}^{\text{out}}}\right\}, \\
&\mathcal{G}^{AC\rightarrow B}(\sigma_{ABC}^{\text{out}})=\max\left\{0,\, \frac{1}{2}\ln \frac{\det AC}{\det\sigma_{ABC}^{\text{out}}}\right\}, \\
&\mathcal{G}^{AB\rightarrow C}(\sigma_{ABC}^{\text{out}})=\max\left\{0,\, \ln\frac{(1+2s^2\sin^2\phi)(1-\theta_k^2)}{1+\theta_k^2(1+2s^2\sin^2\phi)}\right\},
\end{aligned}
\end{eqnarray}
where
\begin{eqnarray*}
\begin{aligned}
&\det BC=\frac{1+2s^2\cos^2\phi+\theta_k^2(1+2s^2)}{1-\theta_k^2}, \\
&\det AC=\frac{1+2s^2+\theta_k^2(1+2s^2\cos^2\phi)}{1-\theta_k^2}, \\
&\det\sigma_{ABC}^{\text{out}}=\left[\frac{1+\theta_k^2(1+2s^2\sin^2\phi)}{1-\theta_k^2}\right]^{2}.
\end{aligned}
\end{eqnarray*}

Using the covariance matrix $\sigma_{ABC}^{\text{out}}$ and Eq.(\ref{m=2}), we derive the analytic expressions of Gaussian steering for all $1\rightarrow2$ partitions of Eq.(\ref{ABC}).

$\textbf{Case(1): $A\rightarrow BC$}$. In this case, the symplectic eigenvalue
$\bar{\nu}_1=\bar{\nu}_2=\left[\frac{1+\theta_k^2(1+2s^2\sin^2\phi)}{(1-\theta_k^2)(1+2s^2\cos^2\phi)}\right]^{\frac{1}{2}}$, which yields the expression
\begin{eqnarray}
\small
\begin{aligned}
\mathcal{G}^{A\rightarrow BC}(\sigma_{ABC}^{\text{out}})=\max\left\{0,\, \ln \frac{(1-\theta_k^2)(1+2s^2\cos^2\phi)}{1+\theta_k^2(1+2s^2\sin^2\phi)}\right\}.
\end{aligned}
\end{eqnarray}

$\textbf{Case(2): $B\rightarrow AC$}$. In this case, the symplectic eigenvalue
$\bar{\nu}_1=\bar{\nu}_2=\left[\frac{1+\theta_k^2(1+2s^2\sin^2\phi)}{(1-\theta_k^2)(1+2s^2)}\right]^{\frac{1}{2}}$, which yields the expression
\begin{eqnarray}
\small
\begin{aligned}
\mathcal{G}^{B\rightarrow AC}(\sigma_{ABC}^{\text{out}})=\max\left\{0,\, \ln \frac{(1-\theta_k^2)(1+2s^2)}{1+\theta_k^2(1+2s^2\sin^2\phi)}\right\}.
\end{aligned}
\end{eqnarray}

$\textbf{Case(3): $C\rightarrow AB$}$. In this case, the symplectic eigenvalue
$\bar{\nu}_1=\bar{\nu}_2=\left[\frac{1+\theta_k^2(1+2s^2\sin^2\phi)}{1+2s^2\sin^2\phi+\theta_k^2}\right]^{\frac{1}{2}}$, which yields the expression
\begin{eqnarray}
\small
\begin{aligned}
\mathcal{G}^{C\rightarrow AB}(\sigma_{ABC}^{\text{out}})=\max\left\{0,\, \ln \frac{1+\theta_k^2(1+2s^2\sin^2\phi)}{\theta_k^2+1+2s^2\sin^2\phi}\right\}.
\end{aligned}
\end{eqnarray}

\begin{figure}[htbp]
\centering
\includegraphics[scale=0.55]{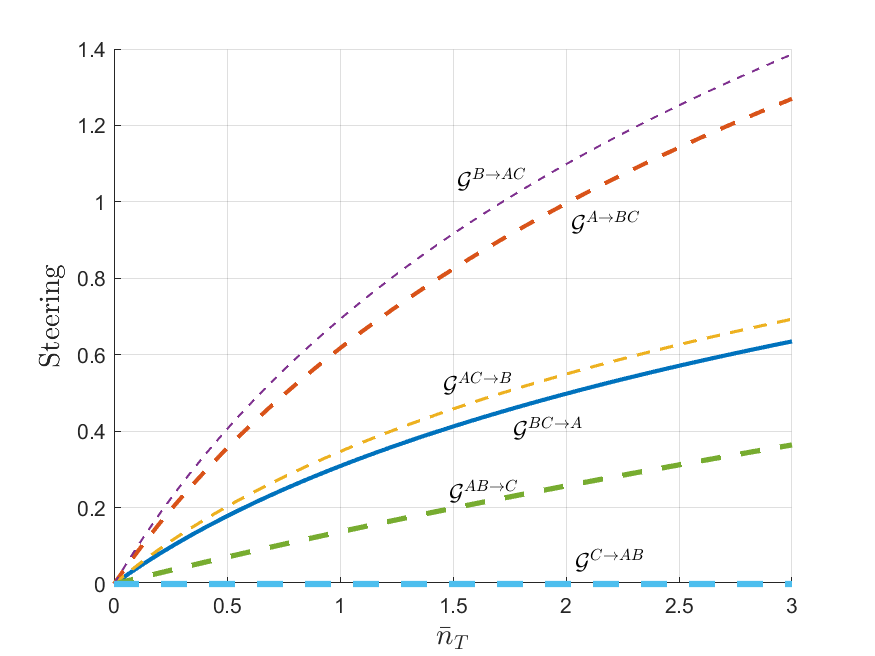}
\caption{The six quantum steering configurations of $\sigma_{ABC}^{\text{out}}$ as functions of the total mean photon number $\bar{n}_T$ with $\phi=\frac{\pi}{8}$ and $k=m=\epsilon=\sigma=1$.}
\label{Fig1}
\end{figure}

\begin{figure}[htbp]
\centering
\includegraphics[scale=0.55]{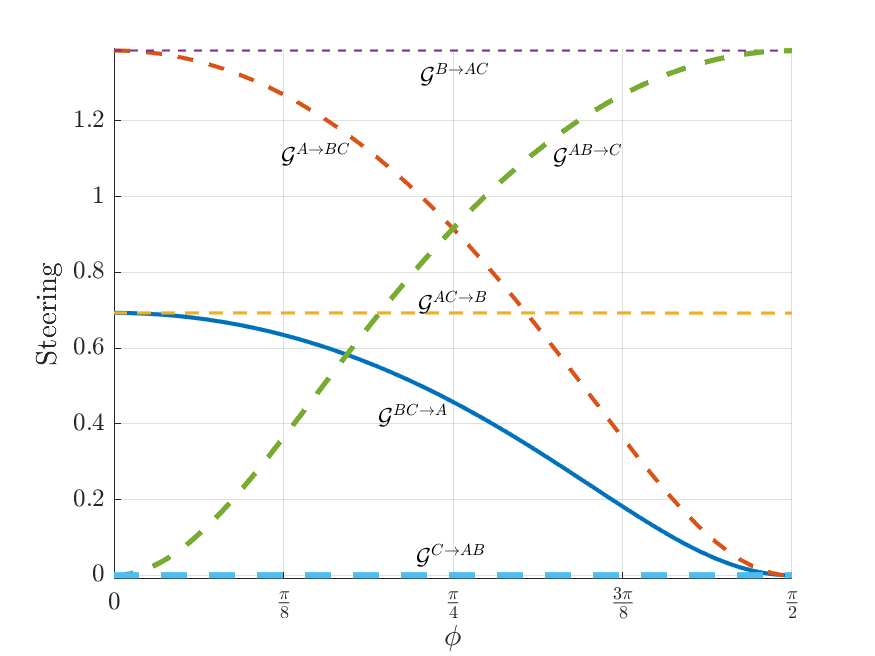}
\caption{The six quantum steering configurations of $\sigma_{ABC}^{\text{out}}$ as functions of $\phi$ with $\bar{n}_T=3$ and $k=m=\epsilon=\sigma=1$.}
\label{Fig2}
\end{figure}

From the above steering formulas, it can be seen that the Gaussian quantum steering of the C3MSV in an expanding spacetime is related to the total mean photon number $\bar{n}_T$, phase parameter $\phi$, and the cosmological expansion parameters (momentum $k$, expanding volume $\epsilon$, mass $m$, expansion rate $\sigma$).

In Fig.~\ref{Fig1}, we plot all $1\rightarrow2$ and $2\rightarrow1$ steering configurations as functions of the total mean photon number $\bar{n}_T$, with fixed $\phi=\frac{\pi}{8}$ and $k=m=\epsilon=\sigma=1$. From Fig.~\ref{Fig1}, one can observe that $\mathcal{G}^{C\rightarrow AB}(\sigma_{ABC}^{\text{out}})=0$ across the entire parameter range. Beyond this case, the remaining five steering configurations monotonically rise as total mean photon number $\bar{n}_T$ increases.
In Fig.~\ref{Fig2}, we depict the six steering configurations as functions of the phase parameter $\phi$, where the parameters are fixed at $\bar{n}_T=3$ and $k=m=\epsilon=\sigma=1$. From Fig.~\ref{Fig2}, it is evident that as the parameter $\phi$ gradually increases, $\mathcal{G}^{A\rightarrow BC}(\sigma_{ABC}^{\text{out}})$ and $\mathcal{G}^{BC\rightarrow A}(\sigma_{ABC}^{\text{out}})$ decrease monotonically, $\mathcal{G}^{AB\rightarrow C}(\sigma_{ABC}^{\text{out}})$ rises monotonically, while the steering channels $\mathcal{G}^{B\rightarrow AC}(\sigma_{ABC}^{\text{out}})$ and $\mathcal{G}^{AC\rightarrow B}(\sigma_{ABC}^{\text{out}})$ remain constant.
From Fig.~\ref{Fig1} and Fig.~\ref{Fig2}, it is clear that the $1\rightarrow2$ steering is not equivalent to that of the $2\rightarrow1$ steering. Furthermore, we can conclude that two-way steering exists between subsystem A and BC, as well as between B and AC, whereas one-way steering emerges between C and AB.

Therefore, we conclude that the steering strength can be amplified by adopting a maximally feasible total mean photon number, while the parameter $\phi$ exerts a dual effect on steerability, capable of either enhancing or suppressing the quantum steering signature.
As demonstrated by He \emph{et al.} in~\cite{He.2013}, our findings corroborate that every individual mode within the C3MSV in the expanding spacetime can be steered by either one or both of the remaining two modes.

\begin{figure*}[htbp]
    \centering
    \begin{minipage}[b]{0.325\textwidth} 
        \includegraphics[width=\linewidth]{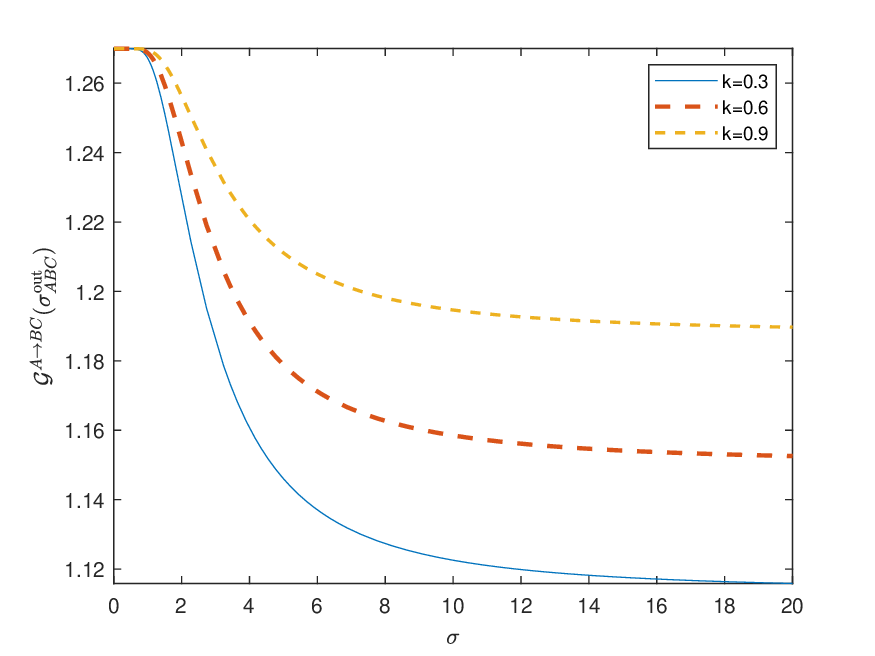} 

    \end{minipage}
    \hfill
    \begin{minipage}[b]{0.325\textwidth}
        \includegraphics[width=\linewidth]{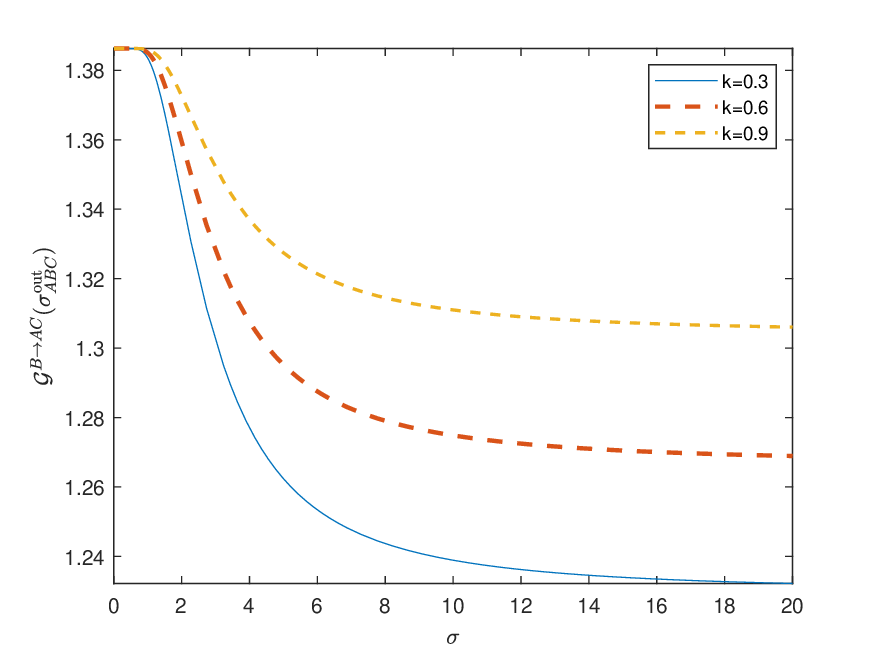}
    \end{minipage}
    \hfill
    \begin{minipage}[b]{0.325\textwidth}
        \includegraphics[width=\linewidth]{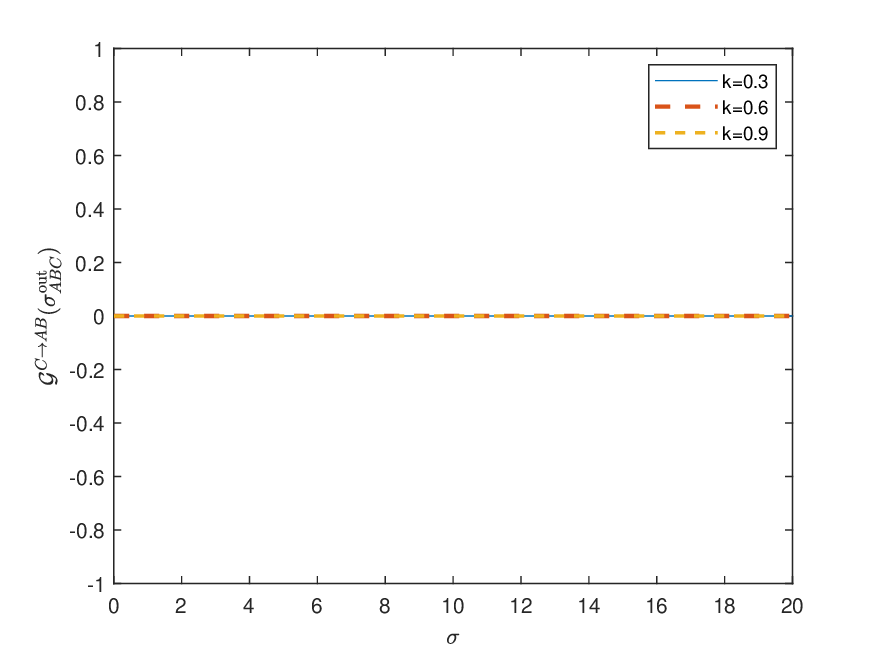}
    \end{minipage}
    \hfill
    \begin{minipage}[b]{0.325\textwidth} 
        \includegraphics[width=\linewidth]{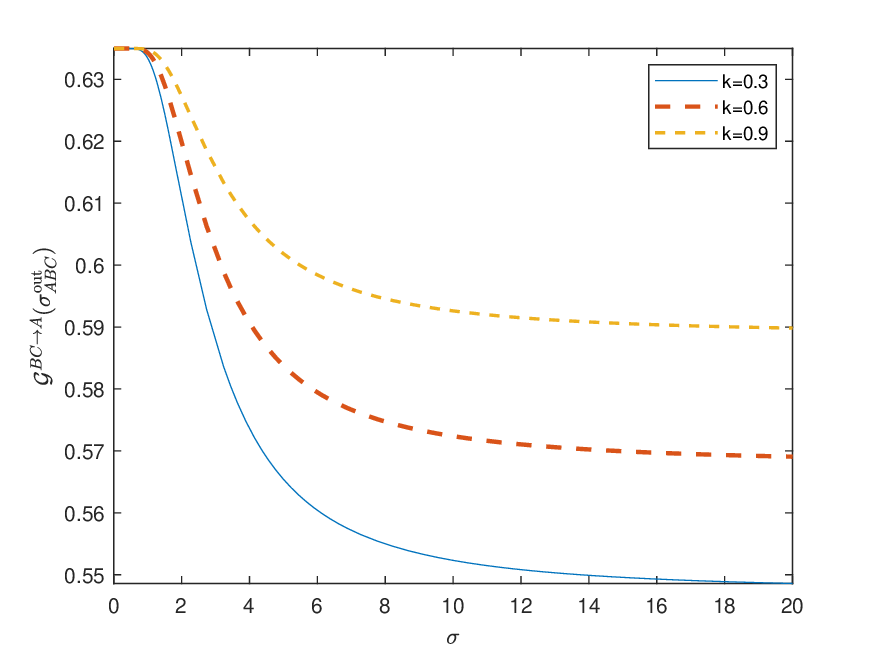} 

    \end{minipage}
    \hfill
    \begin{minipage}[b]{0.325\textwidth}
        \includegraphics[width=\linewidth]{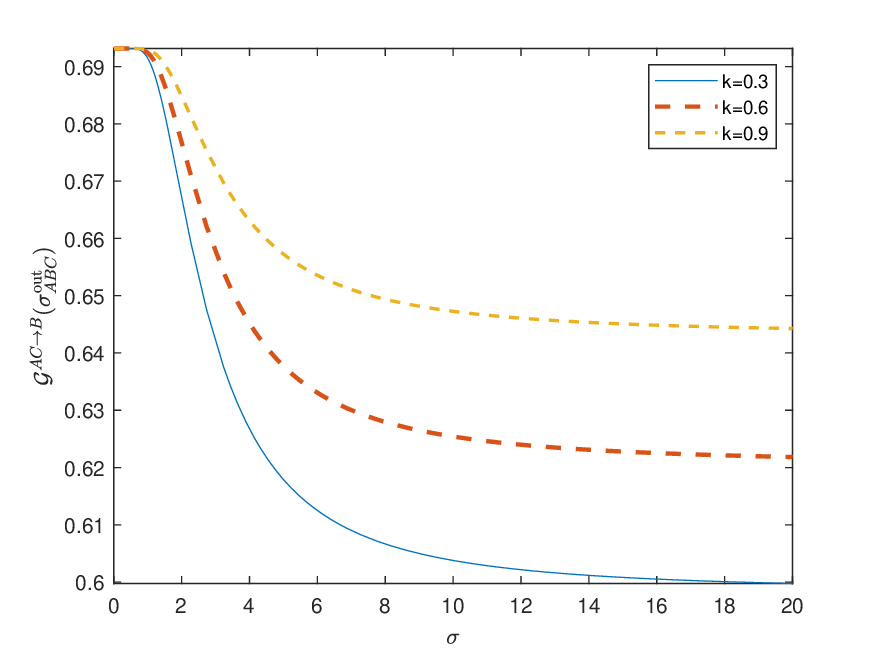}
    \end{minipage}
    \hfill
    \begin{minipage}[b]{0.325\textwidth}
        \includegraphics[width=\linewidth]{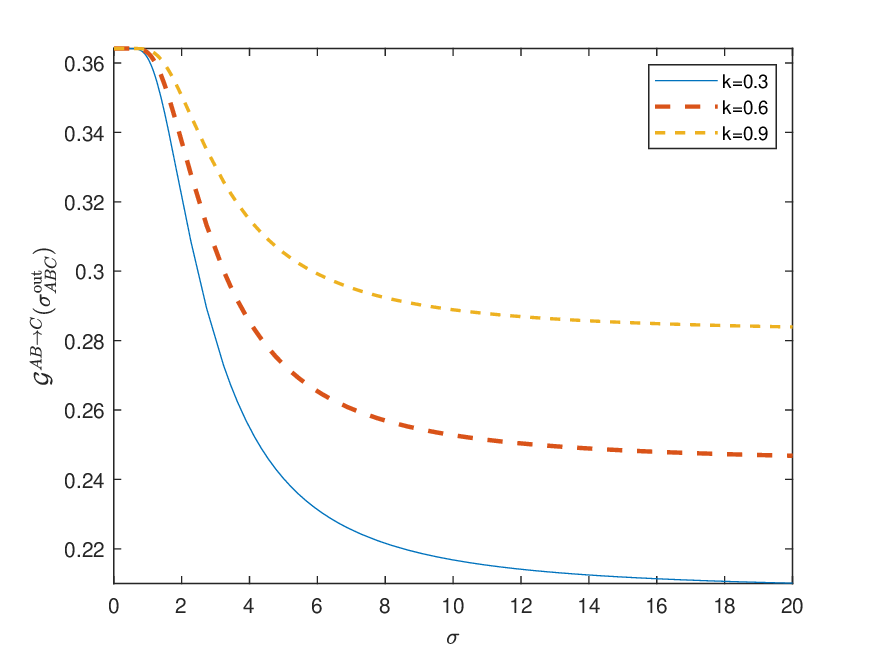}
    \end{minipage}
    \caption{The six quantum steering configurations of $\sigma_{ABC}^{\text{out}}$ as functions of the expansion rate $\sigma$ for different momentum $k$ with $\bar{n}_T=3$, $\phi=\frac{\pi}{8}$ and $m=\epsilon=1$.}
   \label{Fig3}
\end{figure*}

\begin{figure*}[htbp]
    \centering
    \begin{minipage}[b]{0.325\textwidth} 
        \includegraphics[width=\linewidth]{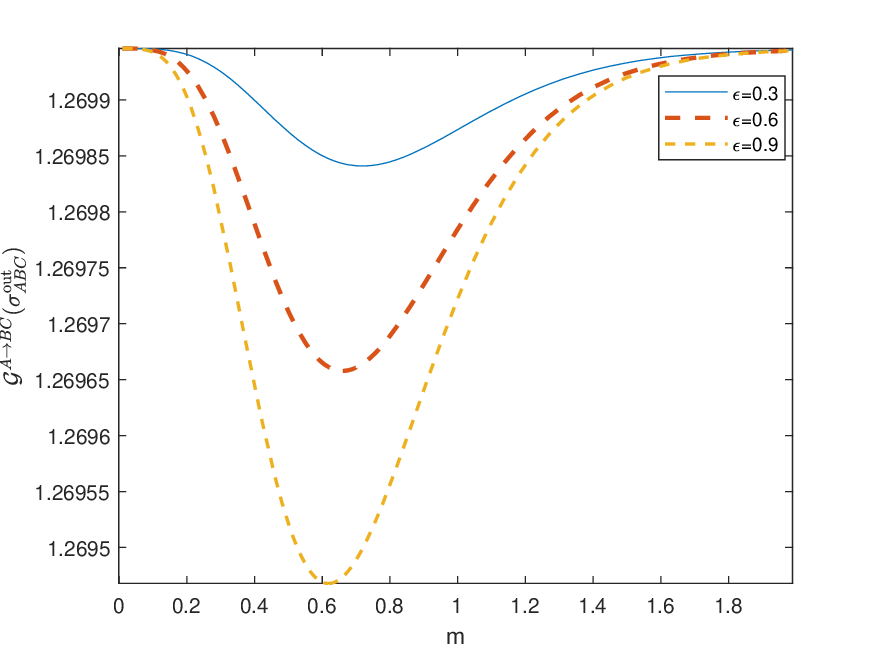} 

    \end{minipage}
    \hfill
    \begin{minipage}[b]{0.325\textwidth}
        \includegraphics[width=\linewidth]{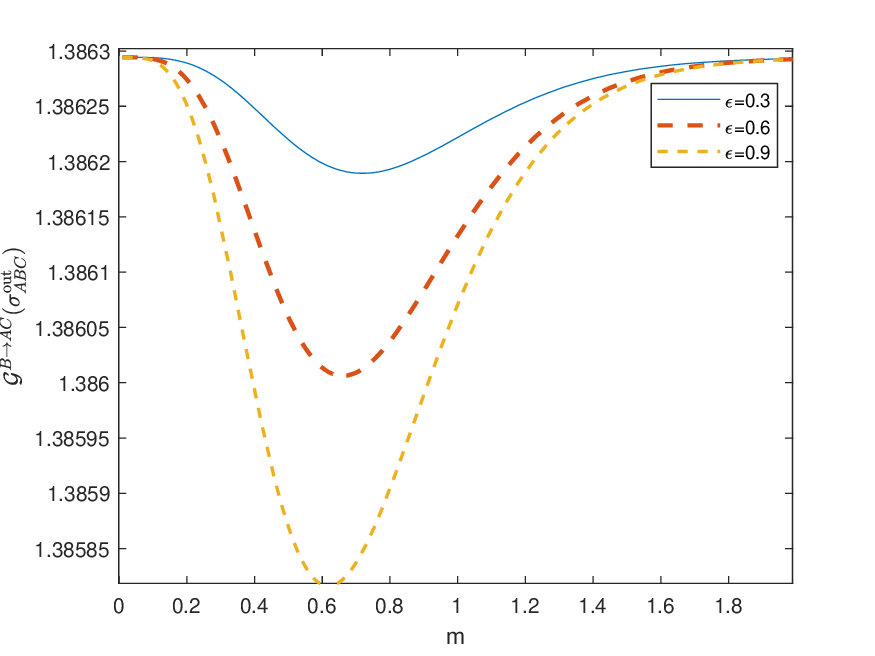}
    \end{minipage}
    \hfill
    \begin{minipage}[b]{0.325\textwidth}
        \includegraphics[width=\linewidth]{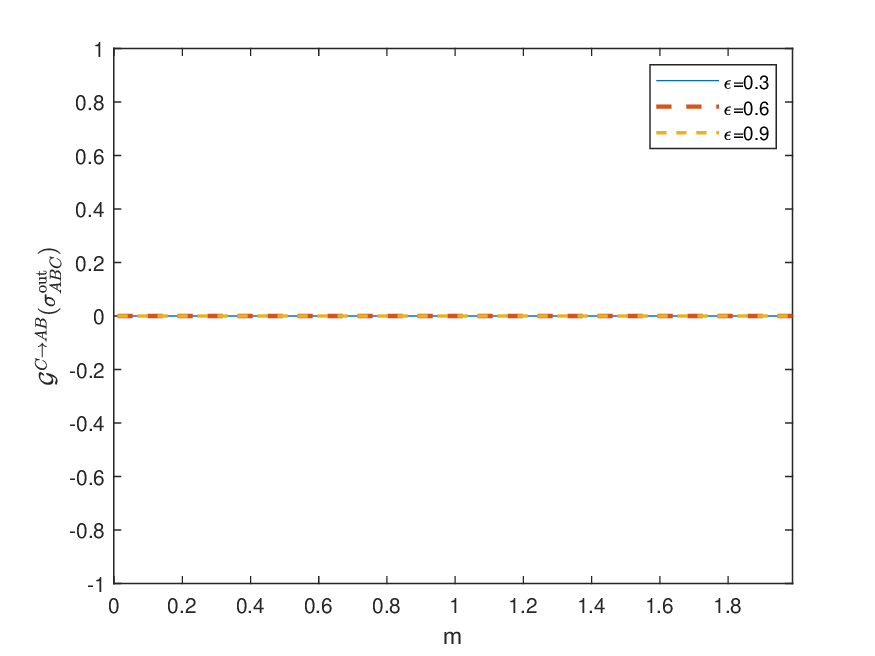}
    \end{minipage}
    \hfill
    \begin{minipage}[b]{0.325\textwidth} 
        \includegraphics[width=\linewidth]{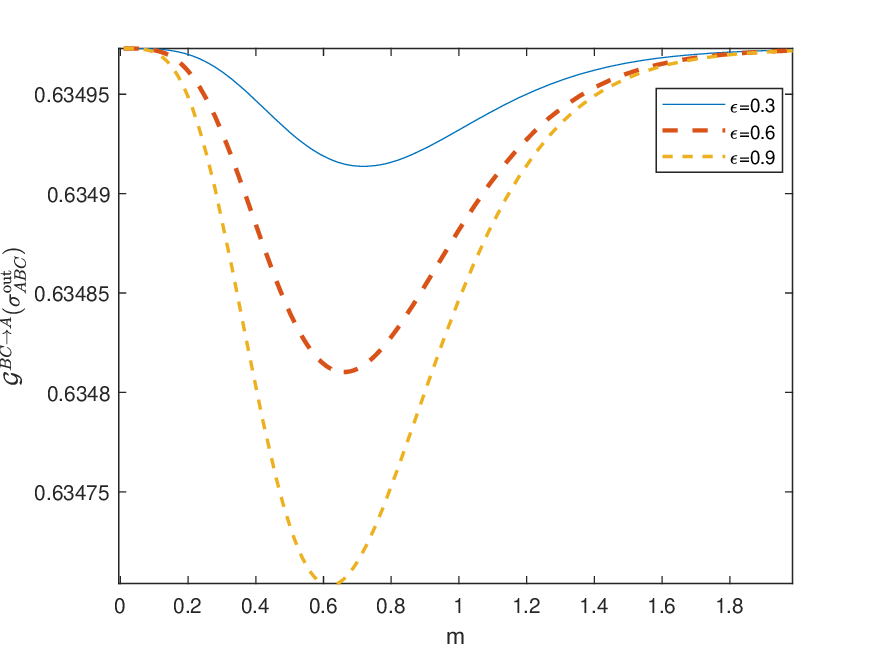} 

    \end{minipage}
    \hfill
    \begin{minipage}[b]{0.325\textwidth}
        \includegraphics[width=\linewidth]{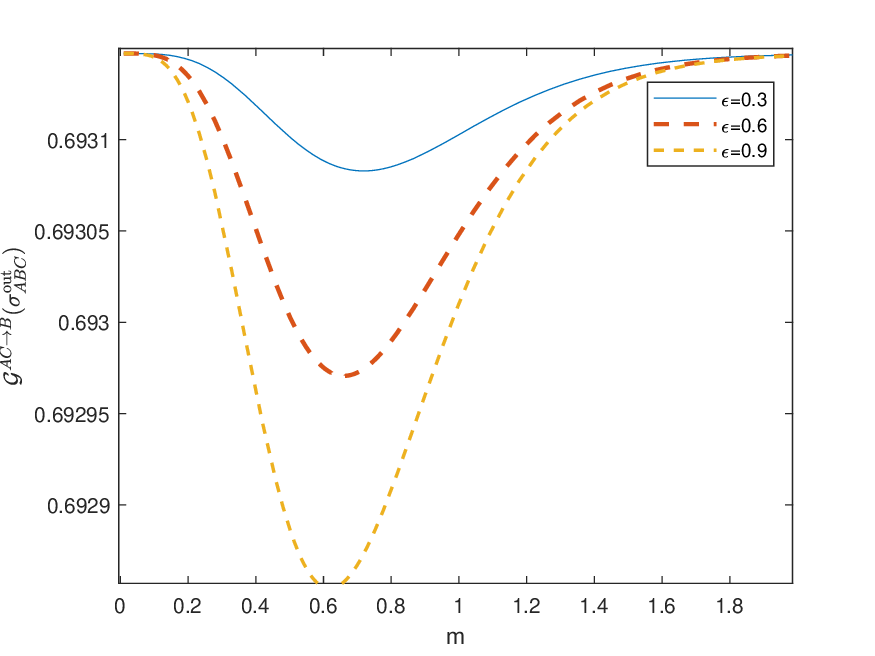}
    \end{minipage}
    \hfill
    \begin{minipage}[b]{0.325\textwidth}
        \includegraphics[width=\linewidth]{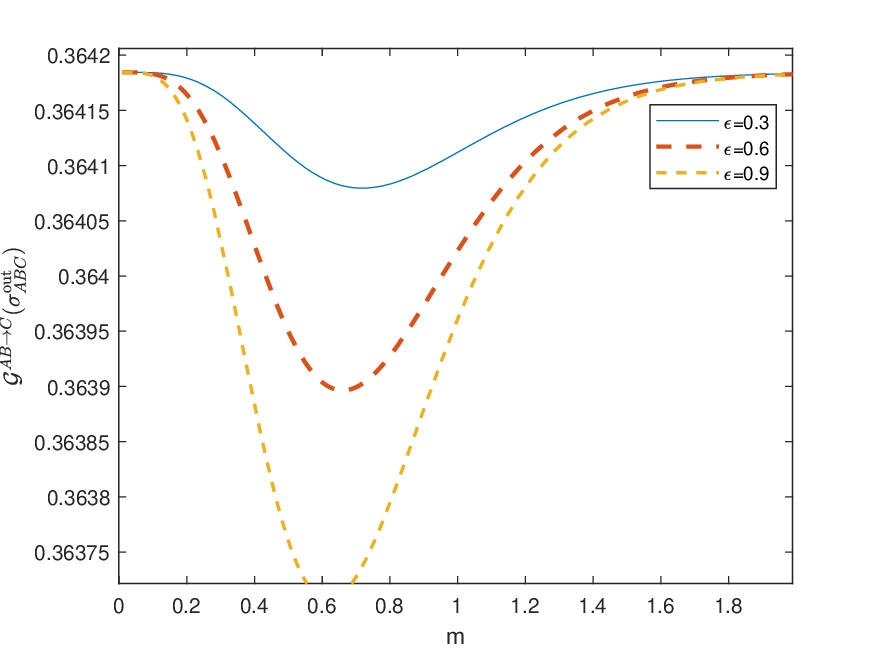}
    \end{minipage}
    \caption{The six quantum steering configurations of $\sigma_{ABC}^{\text{out}}$ as functions of the mass $m$ for different expansion volumes $\epsilon$ with $\bar{n}_T=3$, $\phi=\frac{\pi}{8}$ and $k=\sigma=1$.}
   \label{Fig4}
\end{figure*}

In Fig.~\ref{Fig3} and Fig.~\ref{Fig4}, we plot the six quantum steering configurations versus the expansion rate $\sigma$ (for various momentum $k$) and the mass $m$ (for different expansion volumes $\epsilon$), respectively, with fixed parameters $\bar{n}_T=3, \phi=\frac{\pi}{8}$, $m=\epsilon=1$ in Fig.~\ref{Fig3} and $k=\sigma=1$ in Fig.~\ref{Fig4}. As seen from the Fig.~\ref{Fig3}, in addition to $\mathcal{G}^{C\rightarrow AB}(\sigma_{ABC}^{\text{out}})=0$, the other steering configurations decrease gradually and converge to a steady value as the expansion rate $\sigma$ increases. For a fixed expansion rate $\sigma$, a larger momentum  leads to stronger steering. As illustrated in the Fig.~\ref{Fig4}, the most of the steering configurations first decrease to a minimum and then rise gradually toward a saturated value as the mass $m$ increases. For a fixed mass, a larger expansion volume corresponds to weaker steering. Moreover, quantum steering exhibits strong sensitivity to variations in the mass $m$ around the minimum. In contrast, for values of $m$ far exceeding the minimum point, quantum steering converges to a constant.

Accordingly, we draw the conclusion that greater momentum, combined with reduced expansion volume and expansion rate, leads to stronger quantum steering for the two subsystems. This conclusion is also validated by~\cite{EPJC.2024}.

\subsection{B. Gaussian steering distribution of physically inaccessible modes}

In this subsection, we explore the distribution of Gaussian quantum steering pertaining to physically inaccessible modes.
Accordingly, the covariance matrix describing the two physically accessible modes A, B and the inaccessible mode $\bar{C}$ is obtained via tracing over mode C, with its form given by
\begin{eqnarray}
\label{ABc}
\begin{aligned}
\sigma_{AB\bar{C}}^{\text{out}}=
\begin{pmatrix}
\sigma_A                         & \mathcal{E}_{AB}                & \mathcal{E}_{A\bar{C}}             \\
\mathcal{E}_{AB}^T               & \sigma_B                        & \mathcal{E}_{B\bar{C}}             \\
\mathcal{E}_{A\bar{C}}^T         & \mathcal{E}_{B\bar{C}}^T        & \sigma_{\bar{C}}                     \\
\end{pmatrix}.
\end{aligned}
\end{eqnarray}

With the covariance matrix $\sigma_{AB\bar{C}}^{\text{out}}$ and Eq.(\ref{m=1}) at hand, we solve for the analytical formulas of Gaussian steering across all $2\rightarrow1$ partitions of the state from Eq.(\ref{ABc}). The derived steering are written as
\begin{eqnarray}
\begin{aligned}
&\mathcal{G}^{B\bar{C}\rightarrow A}(\sigma_{AB\bar{C}}^{\text{out}})=\max\left\{0,\, \frac{1}{2}\ln \frac{\det B\bar{C}}{\det\sigma_{AB\bar{C}}^{\text{out}}}\right\}, \\
&\mathcal{G}^{A\bar{C}\rightarrow B}(\sigma_{AB\bar{C}}^{\text{out}})=\max\left\{0,\, \frac{1}{2}\ln \frac{\det A\bar{C}}{\det\sigma_{AB\bar{C}}^{\text{out}}}\right\}, \\
&\mathcal{G}^{AB\rightarrow \bar{C}}(\sigma_{AB\bar{C}}^{\text{out}})=\max\left\{0,\, \ln\frac{(1+2s^2\sin^2\phi)(1-\theta_k^2)}{\theta_k^2+1+2s^2\sin^2\phi}\right\},
\end{aligned}
\end{eqnarray}
where
\begin{eqnarray*}
\begin{aligned}
&\det B\bar{C}=\frac{1+2s^2+\theta_k^2(1+2s^2\cos^2\phi)}{1-\theta_k^2}, \\
&\det A\bar{C}=\frac{1+2s^2\cos^2\phi+\theta_k^2(1+2s^2)}{1-\theta_k^2}, \\
&\det\sigma_{AB\bar{C}}^{\text{out}}=\left(\frac{\theta_k^2+1+2s^2\sin^2\phi}{1-\theta_k^2}\right)^{2}.
\end{aligned}
\end{eqnarray*}

Utilizing the covariance matrix $\sigma_{AB\bar{C}}^{\text{out}}$ and Eq.(\ref{m=2}), we obtain the analytical formulas for Gaussian steering over all $1\rightarrow2$ partitions of the quantum state given in Eq.(\ref{ABc}).

$\textbf{Case(1): $A\rightarrow B\bar{C}$}$. In this case, the symplectic eigenvalue
$\bar{\nu}_1=\bar{\nu}_2=\left[\frac{\theta_k^2+1+2s^2\sin^2\phi}{(1-\theta_k^2)(1+2s^2\cos^2\phi)}\right]^{\frac{1}{2}}$, which leads to
\begin{eqnarray}
\small
\begin{aligned}
\mathcal{G}^{A\rightarrow B\bar{C}}(\sigma_{AB\bar{C}}^{\text{out}})=\max\left\{0,\, \ln \frac{(1-\theta_k^2)(1+2s^2\cos^2\phi)}{\theta_k^2+1+2s^2\sin^2\phi}\right\}.
\end{aligned}
\end{eqnarray}

$\textbf{Case(2): $B\rightarrow A\bar{C}$}$. In this case, the symplectic eigenvalue
$\bar{\nu}_1=\bar{\nu}_2=\left[\frac{\theta_k^2+1+2s^2\sin^2\phi}{(1-\theta_k^2)(1+2s^2)}\right]^{\frac{1}{2}}$, which which leads to
\begin{eqnarray}
\small
\begin{aligned}
\mathcal{G}^{B\rightarrow A\bar{C}}(\sigma_{AB\bar{C}}^{\text{out}})=\max\left\{0,\, \ln \frac{(1-\theta_k^2)(1+2s^2)}{\theta_k^2+1+2s^2\sin^2\phi}\right\}.
\end{aligned}
\end{eqnarray}

$\textbf{Case(3): $\bar{C}\rightarrow AB$}$. In this case, the symplectic eigenvalue
$\bar{\nu}_1=\bar{\nu}_2=\left[\frac{1+\theta_k^2(1+2s^2\sin^2\phi)}{1+2s^2\sin^2\phi+\theta_k^2}\right]^{\frac{1}{2}}$, which which leads to
\begin{eqnarray}
\small
\begin{aligned}
\mathcal{G}^{B\rightarrow A\bar{C}}(\sigma_{AB\bar{C}}^{\text{out}})=\max\left\{0,\, \ln \frac{1+2s^2\sin^2\phi+\theta_k^2}{1+\theta_k^2(1+2s^2\sin^2\phi)}\right\}.
\end{aligned}
\end{eqnarray}

\begin{figure}[htbp]
\centering
\includegraphics[scale=0.55]{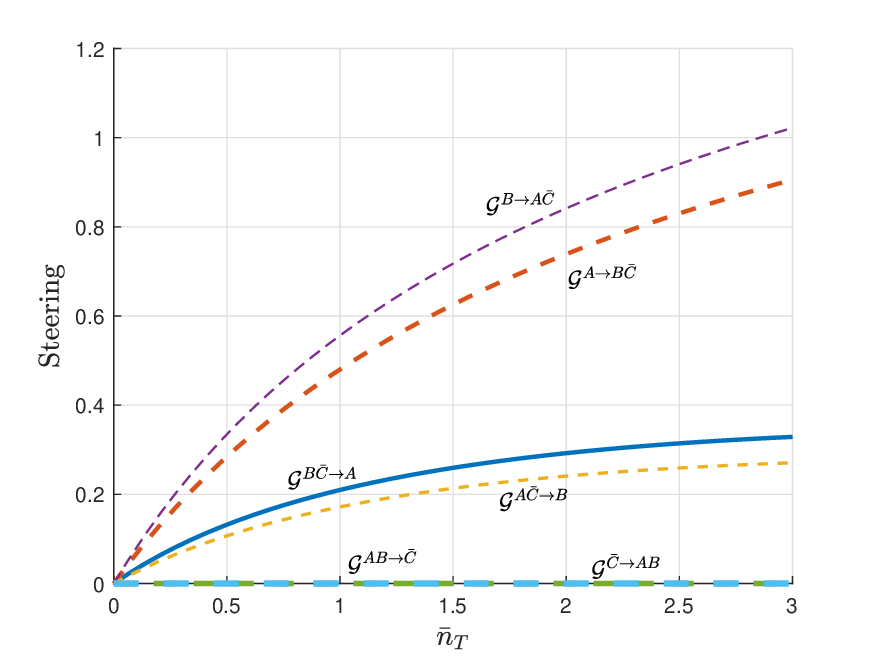}
\caption{The six quantum steering configurations of $\sigma_{AB\bar{C}}^{\text{out}}$ as functions of the total mean photon number $\bar{n}_T$ with $\phi=\frac{\pi}{8}$ and $k=m=\epsilon=\sigma=1$.}
\label{Fig5}
\end{figure}

\begin{figure}[htbp]
\centering
\includegraphics[scale=0.55]{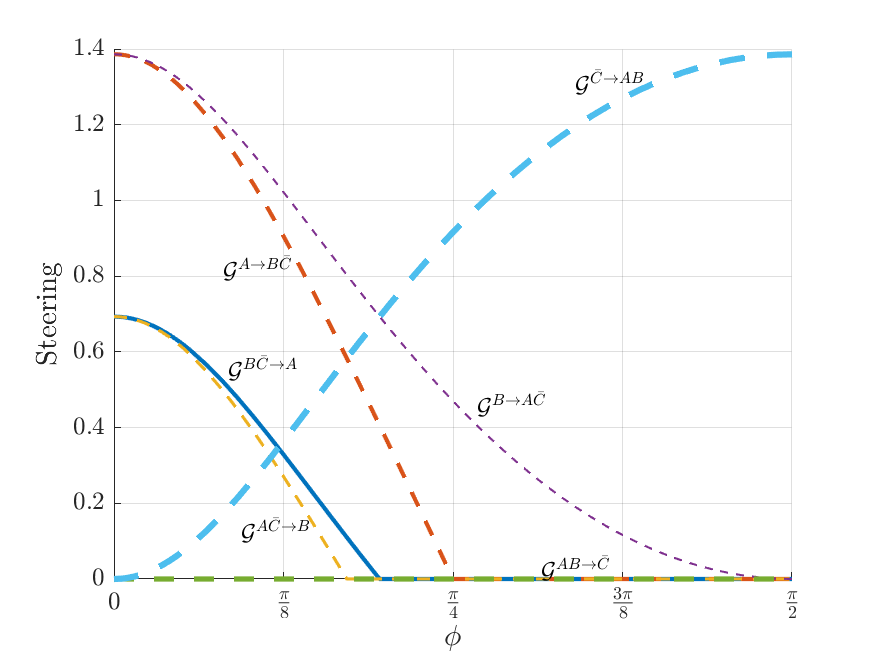}
\caption{The six quantum steering configurations of $\sigma_{AB\bar{C}}^{\text{out}}$ as functions of $\phi$ with $\bar{n}_T=3$ and $k=m=\epsilon=\sigma=1$.}
\label{Fig6}
\end{figure}

In Fig.~\ref{Fig5}, we present all $1\rightarrow2$ and $2\rightarrow1$ steering configurations versus the total mean photon number $\bar{n}_T$, where $\phi=\frac{\pi}{8}$ and $k=m=\epsilon=\sigma=1$ are kept constant. It can be seen from Fig.~\ref{Fig5} that $\mathcal{G}^{\bar{C}\rightarrow AB}(\sigma_{AB\bar{C}}^{\text{out}})=\mathcal{G}^{AB\rightarrow \bar{C}}(\sigma_{AB\bar{C}}^{\text{out}})=0$ for all considered parameters. All other steering configurations increase monotonically with growing $\bar{n}_T$. Furthermore, two-way steering exists between subsystem A and BC, as well as between B and AC. By contrast, no-way steering emerges between subsystem C and AB.

In Fig.~\ref{Fig6}, we illustrate the six steering configurations versus the phase parameter $\phi$, with $\bar{n}_T=3$ and $k=m=\epsilon=\sigma=1$ held constant.
Fig.~\ref{Fig6} clearly demonstrates that four quantities $\mathcal{G}^{A\rightarrow B\bar{C}}(\sigma_{AB\bar{C}}^{\text{out}})$, $\mathcal{G}^{B\bar{C}\rightarrow A}(\sigma_{AB\bar{C}}^{\text{out}})$, $\mathcal{G}^{B\rightarrow A\bar{C}}(\sigma_{AB\bar{C}}^{\text{out}})$ and $\mathcal{G}^{A\bar{C}\rightarrow B}(\sigma_{AB\bar{C}}^{\text{out}})$ undergo a monotonic reduction when $\phi$ goes up slowly. Conversely, $\mathcal{G}^{\bar{C}\rightarrow AB}(\sigma_{AB\bar{C}}^{\text{out}})$ presents a monotonic upward trend. Notably, the steering $\mathcal{G}^{A\rightarrow B\bar{C}}(\sigma_{AB\bar{C}}^{\text{out}})$, $\mathcal{G}^{B\bar{C}\rightarrow A}(\sigma_{AB\bar{C}}^{\text{out}})$ and $\mathcal{G}^{A\bar{C}\rightarrow B}(\sigma_{AB\bar{C}}^{\text{out}})$ exhibits a ``sudden death" phenomenon, whereas $\mathcal{G}^{B\rightarrow A\bar{C}}(\sigma_{AB\bar{C}}^{\text{out}})$ gradually decays to zero as $\phi$ increases.
As $\phi$ increases, the steering between subsystem $A$ and $B\bar{C}$ evolves successively from two-way to one-way, and finally from one-way to no-way. The transition from two-way to one-way steering coincides with the ``sudden death" of $\mathcal{G}^{B\bar{C}\rightarrow A}(\sigma_{AB\bar{C}}^{\text{out}})$, while the shift toward no-way steering corresponds to the ``sudden death" of $\mathcal{G}^{A\rightarrow B\bar{C}}(\sigma_{AB\bar{C}}^{\text{out}})$.
For subsystems $B$ and $A\bar{C}$, the steering changes from two-way to one-way as $\phi$ increases, which corresponds to the ``sudden death" of $\mathcal{G}^{A\bar{C}\rightarrow B}(\sigma_{AB\bar{C}}^{\text{out}})$.
Meanwhile, the steering between $\bar{C}$ and $AB$ remains one-way throughout.
\begin{figure*}[htbp]
    \centering
    \begin{minipage}[b]{0.325\textwidth} 
        \includegraphics[width=\linewidth]{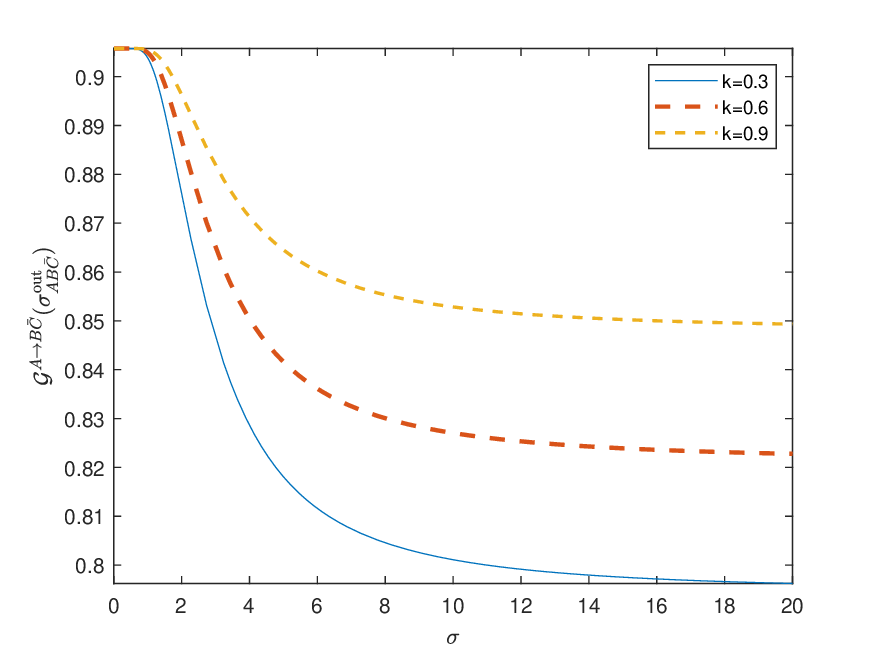} 

    \end{minipage}
    \hfill
    \begin{minipage}[b]{0.325\textwidth}
        \includegraphics[width=\linewidth]{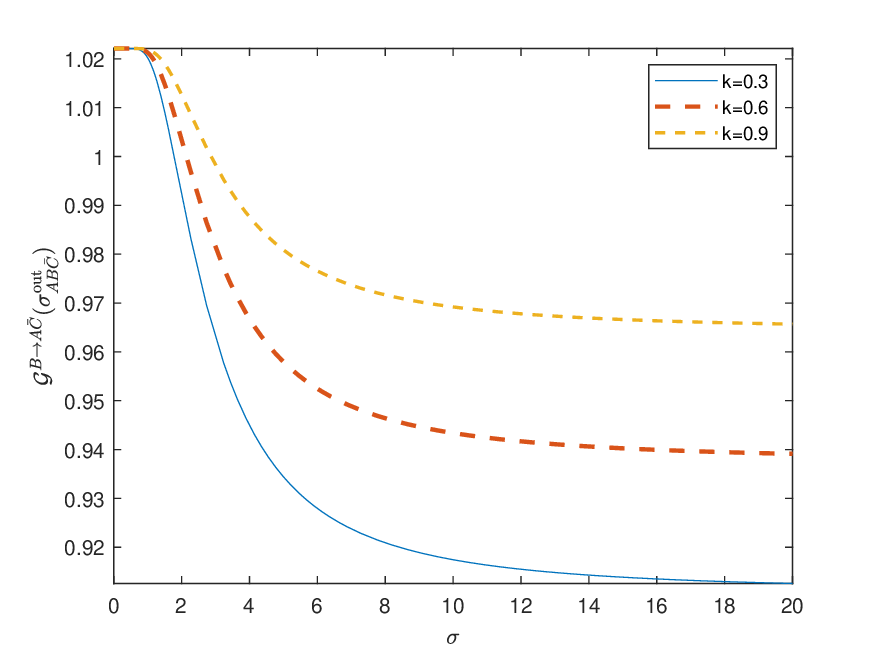}
    \end{minipage}
    \hfill
    \begin{minipage}[b]{0.325\textwidth}
        \includegraphics[width=\linewidth]{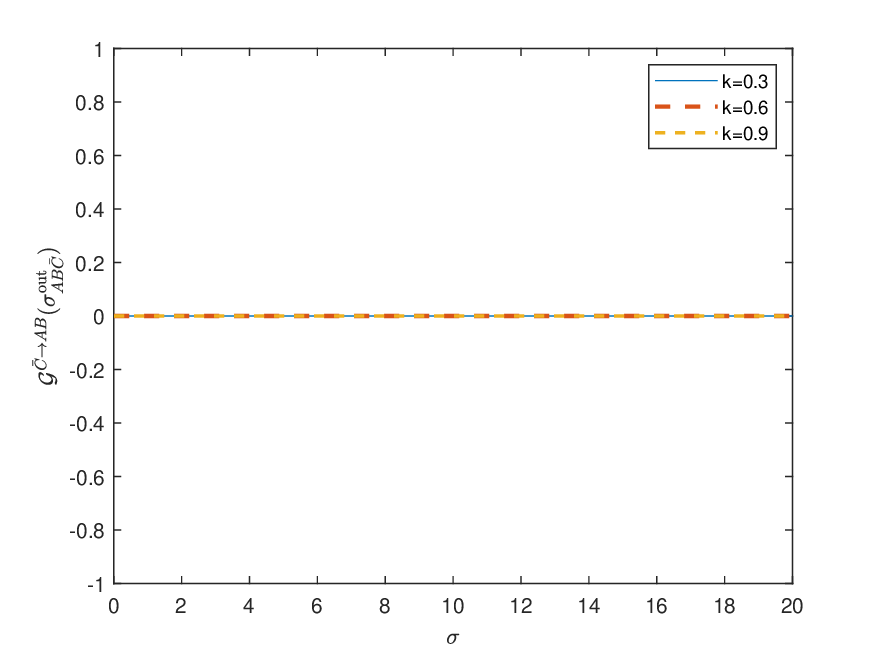}
    \end{minipage}
    \hfill
    \begin{minipage}[b]{0.325\textwidth} 
        \includegraphics[width=\linewidth]{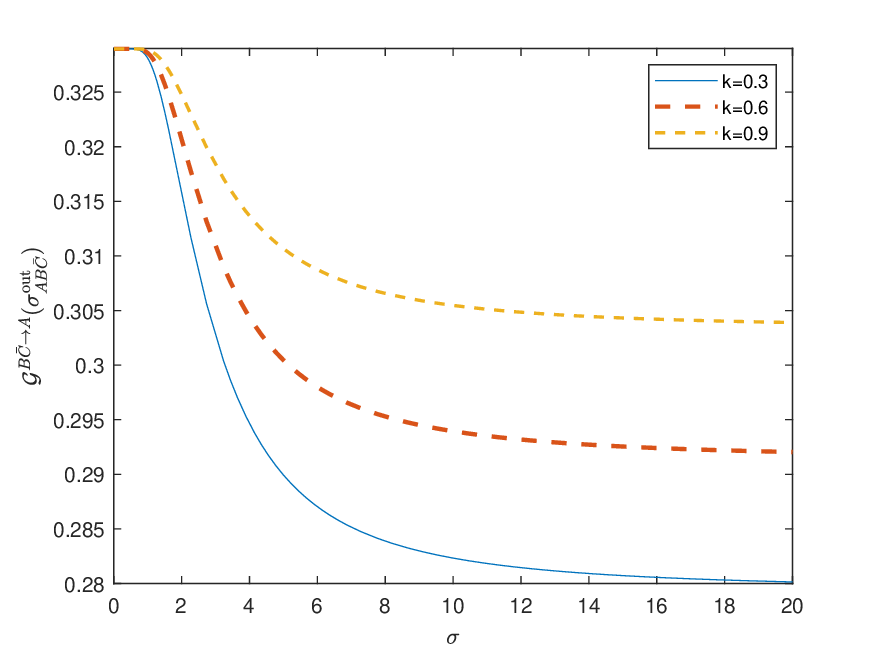} 

    \end{minipage}
    \hfill
    \begin{minipage}[b]{0.325\textwidth}
        \includegraphics[width=\linewidth]{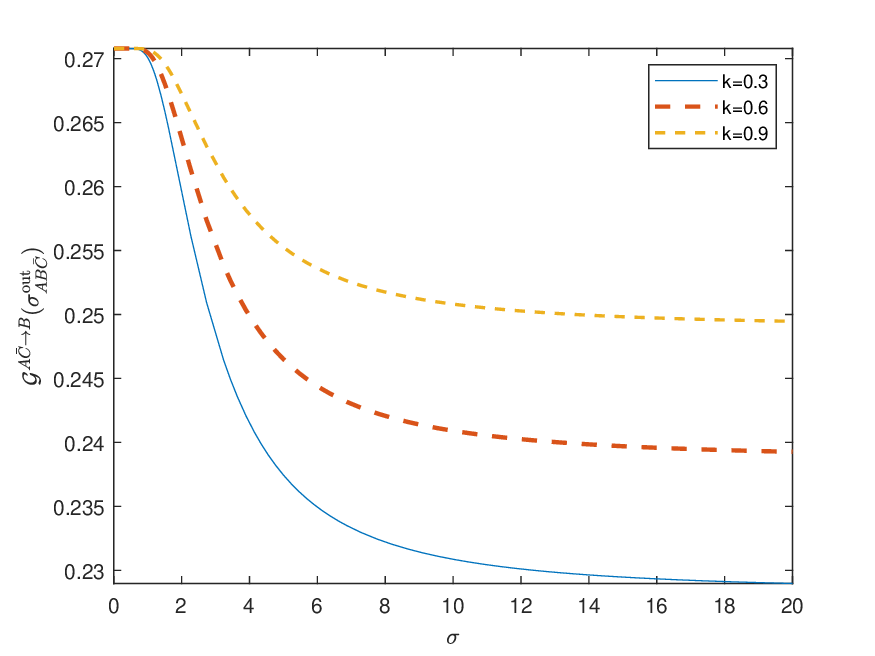}
    \end{minipage}
    \hfill
    \begin{minipage}[b]{0.325\textwidth}
        \includegraphics[width=\linewidth]{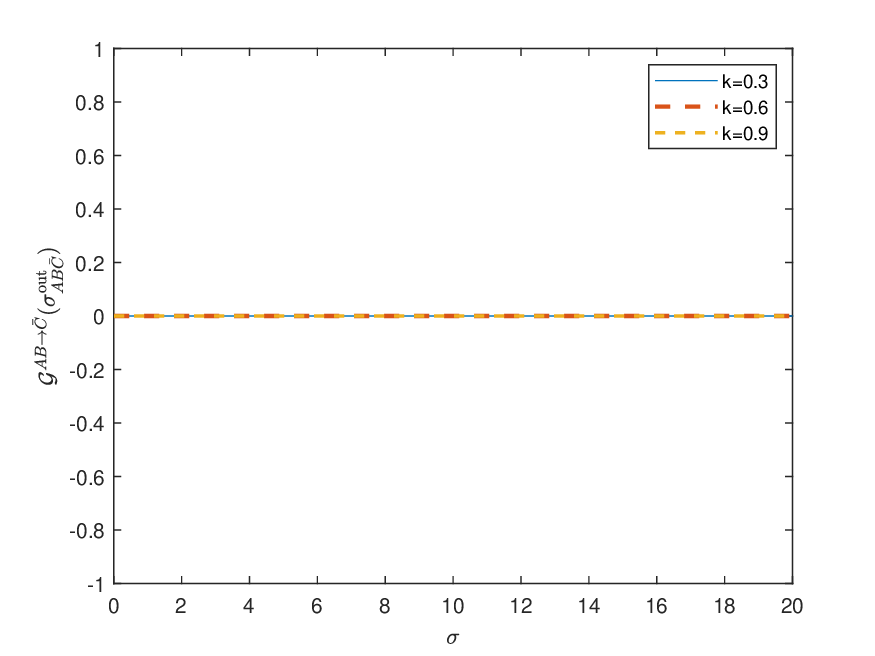}
    \end{minipage}
    \caption{The six quantum steering configurations of $\sigma_{AB\bar{C}}^{\text{out}}$ as functions of the expansion rate $\sigma$ for different momentum $k$ with $\bar{n}_T=3$, $\phi=\frac{\pi}{8}$ and $m=\epsilon=1$.}
   \label{Fig7}
\end{figure*}

\begin{figure*}[htbp]
    \centering
    \begin{minipage}[b]{0.325\textwidth} 
        \includegraphics[width=\linewidth]{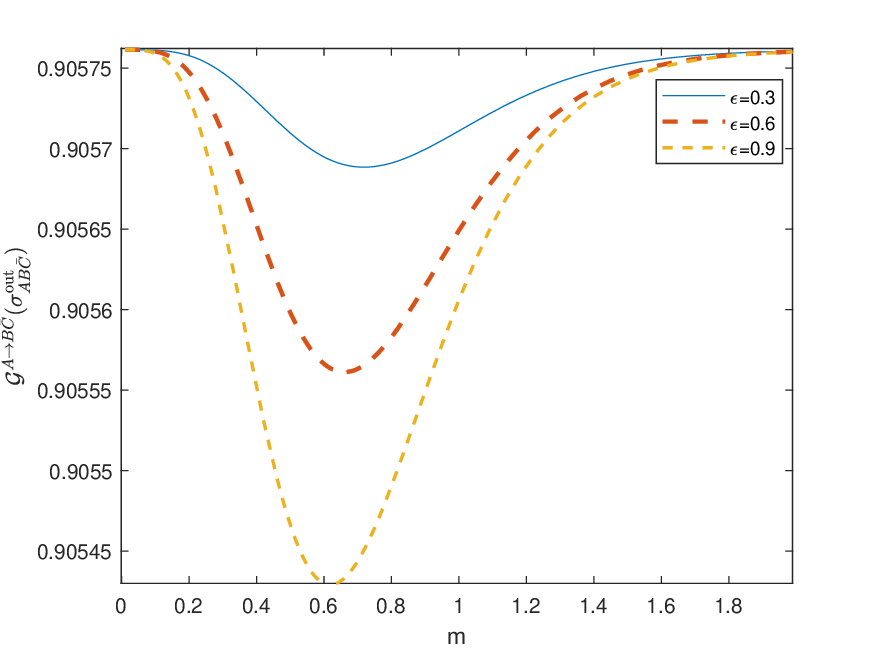} 

    \end{minipage}
    \hfill
    \begin{minipage}[b]{0.325\textwidth}
        \includegraphics[width=\linewidth]{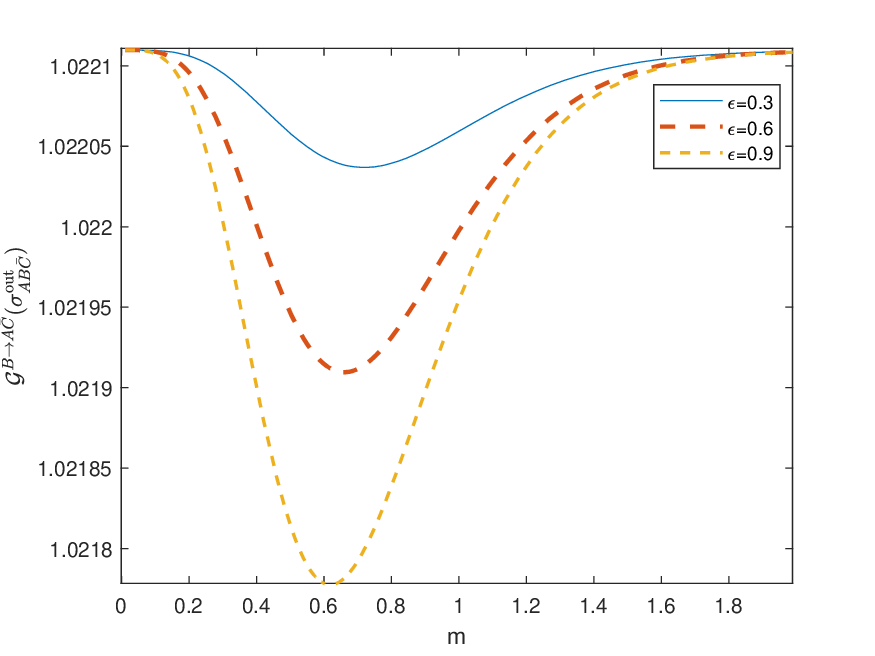}
    \end{minipage}
    \hfill
    \begin{minipage}[b]{0.325\textwidth}
        \includegraphics[width=\linewidth]{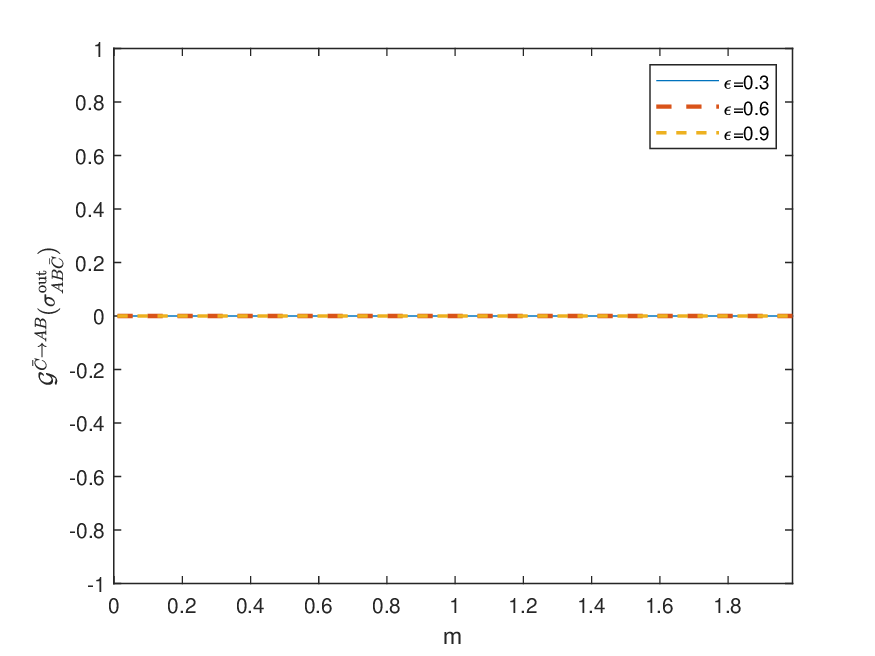}
    \end{minipage}
    \hfill
    \begin{minipage}[b]{0.325\textwidth} 
        \includegraphics[width=\linewidth]{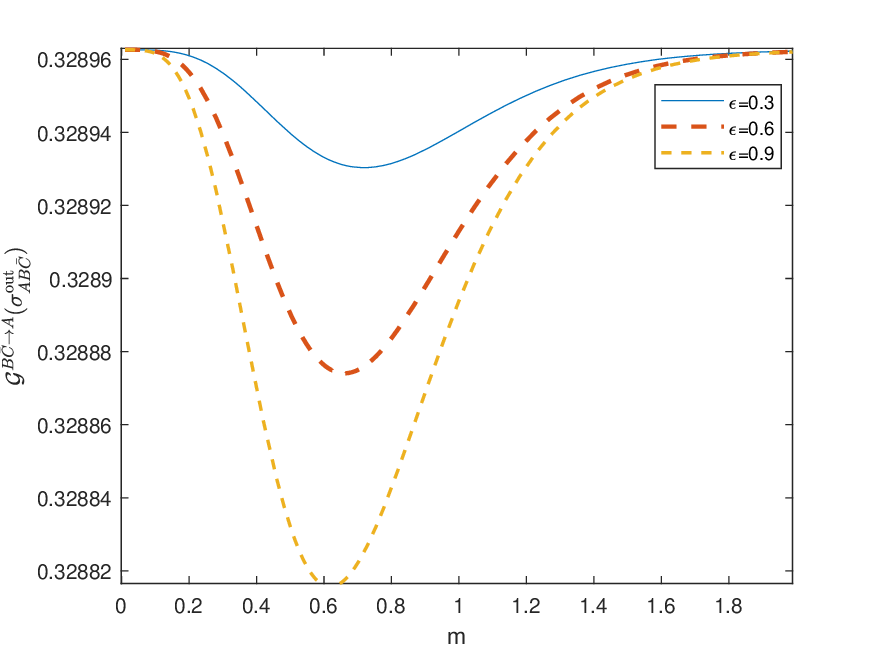} 

    \end{minipage}
    \hfill
    \begin{minipage}[b]{0.325\textwidth}
        \includegraphics[width=\linewidth]{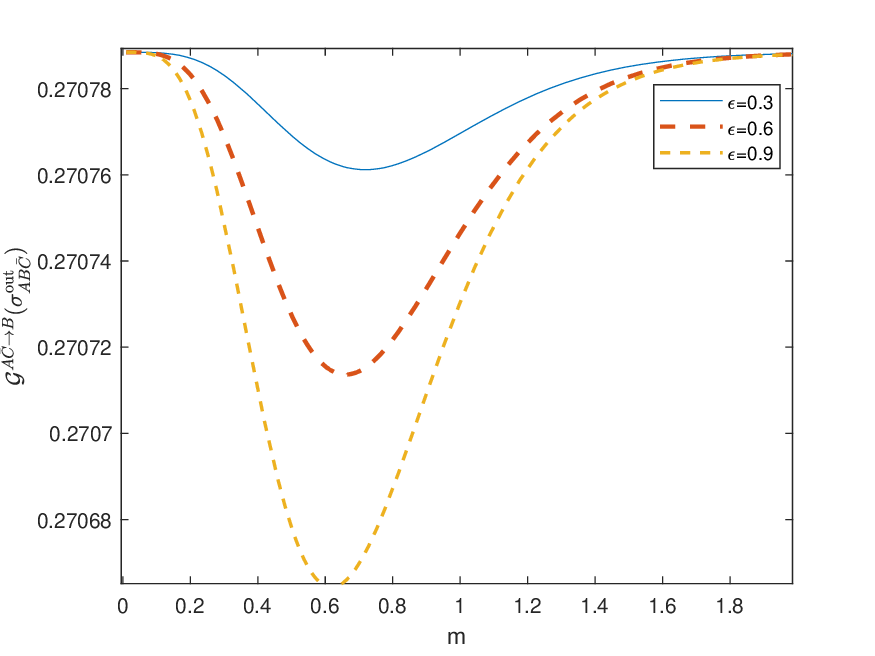}
    \end{minipage}
    \hfill
    \begin{minipage}[b]{0.325\textwidth}
        \includegraphics[width=\linewidth]{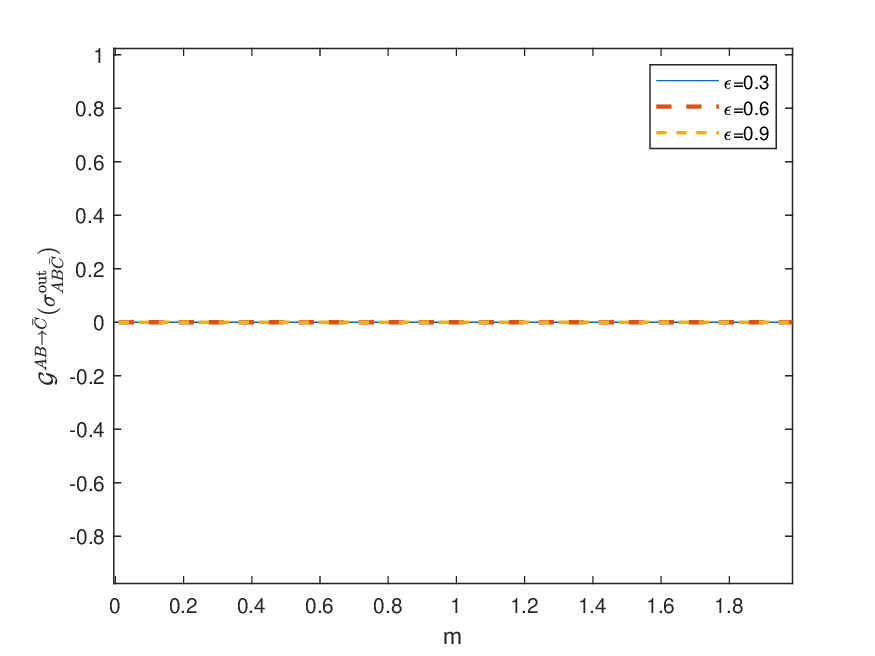}
    \end{minipage}
    \caption{The six quantum steering configurations of $\sigma_{AB\bar{C}}^{\text{out}}$ as functions of the mass $m$ for different expansion volumes $\epsilon$ with $\bar{n}_T=3$, $\phi=\frac{\pi}{8}$ and $k=\sigma=1$.}
   \label{Fig8}
\end{figure*}

In Fig.~\ref{Fig7} and Fig.~\ref{Fig8}, we present the six quantum steering configurations against the expansion rate $\sigma$ (for different momentum $k$) and mass $m$ (for various expansion volumes $\epsilon$), respectively. The parameters are set as $\bar{n}_T=3, \phi=\frac{\pi}{8}$ and $m=\epsilon=1$ for Fig.~\ref{Fig7}, while $k=\sigma=1$ for Fig.~\ref{Fig8}. As shown in Fig.~\ref{Fig7}, apart from the two quantities satisfying $\mathcal{G}^{\bar{C}\rightarrow AB}(\sigma_{AB\bar{C}}^{\text{out}})=\mathcal{G}^{AB\rightarrow \bar{C}}(\sigma_{AB\bar{C}}^{\text{out}})=0$, the remaining steering configurations
decline slowly and approach a stable value with increasing $\sigma$. At a given $\sigma$, higher momentum yields stronger steering.
Fig.~\ref{Fig8} indicates that most steering configurations first drop to a minimum and then rise steadily toward a saturated value as $m$ grows. For a fixed $m$, a larger expansion volume results in weaker steering. Furthermore, quantum steering is highly sensitive to change in $m$ near the minimum. When $m$ is much larger than this minimum, the steering converges to a constant.

Consequently, higher total mean photon number and momentum, paired with reduced expansion volume and expansion rate, lead to stronger quantum steering of the two subsystems.
In addition, we find that the $1\rightarrow2$ steering is stronger than that $2\rightarrow1$ steering.

\section{Conclusion}
We have investigated the distribution of the Gaussian quantum steering for the C3MSV in an expanding spacetime.
In the paper, we suppose that Alice, Bob and Charlie initially possess the C3MSV $\sigma_{ABC}^{\text{in}}$ in an asymptotically flat spacetime. Afterwards, Charlie experiences cosmic expansion described by the symplectic transformation in Eq.(\ref{symplectic}), whereas Alice and Bob stay in the asymptotically flat domain.
Owing to causal separation between the interior and exterior of the expanding spacetime, Alice, Bob and Charlie are unable to access quantum information beyond the event horizon.
Then, we explore the distribution of Gaussian quantum steering pertaining to physically accessible modes and inaccessible modes, respectively.

Regarding the distribution of Gaussian quantum steering for physically accessible modes, we draw the following conclusions. Steering strength can be enhanced by using the maximally feasible total mean photon number; the parameter $\phi$ exerts a dual effect on steerability, capable of either boosting or weakening quantum steering, and greater momentum combined with reduced expansion volume and expansion rate also strengthens the quantum steering of the two subsystems. As shown in the work of He \emph{et al.}~\cite{He.2013}, our results confirm that each mode of the C3MSV state in the expanding spacetime is steerable by one or both of the other two modes.

For the Gaussian quantum steering distribution of physically inaccessible modes, we conclude that larger total mean photon number and momentum, alongside smaller expansion volume and expansion rate, yield stronger quantum steering.
We also compare the Gaussian steering of the C3MSV state in expanding spacetime investigated here with relevant results from previous studies. It has been reported that no ``sudden death'' occurs for C3MSV Gaussian steering in the flat spacetime~\cite{Zhan.2023} and for bipartite Gaussian steering under an expanding spacetime~\cite{EPJC.2024}. Our results reveal that as $\phi$ increases, the steering $\mathcal{G}^{A\rightarrow B\bar{C}}(\sigma_{AB\bar{C}}^{\text{out}})$, $\mathcal{G}^{B\bar{C}\rightarrow A}(\sigma_{AB\bar{C}}^{\text{out}})$ and $\mathcal{G}^{A\bar{C}\rightarrow B}(\sigma_{AB\bar{C}}^{\text{out}})$ exhibit the ``sudden death'' phenomenon, whereas this behavior is absent in other steering configurations. Besides, the strength of $1\rightarrow2$ steering is always greater than that of $2\rightarrow1$ steering.

Our work substantially enriches the theoretical framework for tripartite quantum steering in expanding spacetime. These results are of significance to the research on quantum steering in curved spacetime, and offer new perspectives and theoretical bases for further investigations.

\begin{acknowledgments}
This work is supported by the National Natural Science Foundation of China (NSFC) under Grant No.12564048; the Natural Science Foundation of Hainan Province under Grant No. 125RC744 and the Key Laboratory of Data Science and Intelligence Education, Ministry of Education, Hainan Normal University.
\end{acknowledgments}



\end{document}